\documentclass[letterpaper,11pt]{article}
\usepackage[width=15cm,height=23cm]{geometry} 
\usepackage[usenames]{color}

\usepackage{amsfonts,amssymb}
\usepackage{theorem}
\usepackage[dvips]{graphicx}

\def\G{\Gamma}

\long\def\symbolfootnote[#1]#2{\begingroup%
\def\thefootnote{\fnsymbol{footnote}}\footnote[#1]{#2}\endgroup}

\newcommand{\slsh}[1]
{
#1 \!\!\!\!\slash
}

\def\slshp{p\!\!\!\slash}

\begin{document}

\par\noindent
\rightline{IFUM-931-FT}
\rightline{MIT-CTP-4009}
\rightline{arXiv:0809.1994}
\rightline{November 2010}


\vskip 1.3 truecm
\bf
\Large
\centerline{The $SU(2) \otimes U(1)$ Electroweak Model based on}
\centerline{the Nonlinearly Realized Gauge Group. II.}
\vskip 0.3 truecm
\large
\centerline{Functional Equations and the Weak Power-Counting
\symbolfootnote[2]{\tt This work is supported in part by funds provided by the U.S. Department
of Energy (D.O.E.) under cooperative research agreement \#DE FG02-05ER41360}
}

\normalsize
\rm

\large
\rm
\vskip 0.7 truecm
\centerline{D.~Bettinelli$^b$\footnote{e-mail: 
{\tt daniele.bettinelli@mi.infn.it}}, 
R.~Ferrari$^{a,b}$\footnote{e-mail: {\tt ruggero.ferrari@mi.infn.it}}, 
A.~Quadri$^b$\footnote{e-mail: {\tt andrea.quadri@mi.infn.it}}}

\normalsize
\medskip
\begin{center}
$^a$
Center for Theoretical Physics\\
Laboratory for Nuclear Science\\
and Department of Physics\\
Massachussetts Institute of Technology\\
Cambridge, Massachussetts 02139
and\\
$^b$
Dip. di Fisica, Universit\`a degli Studi di Milano\\
and INFN, Sez. di Milano\\
via Celoria 16, I-20133 Milano, Italy
\end{center}

%
%
\begin{abstract}
In the present paper, that is the second part devoted
to the construction of an electroweak model based on a nonlinear
realization of the gauge group $SU(2)\otimes U(1)$, we 
study the
tree-level vertex functional
with all the sources necessary 
for the functional formulation of the relevant symmetries
(Local Functional Equation, Slavnov-Taylor identity,
Landau Gauge Equation)
 and for
the symmetric removal of the divergences. The Weak Power Counting
criterion is proven in the presence of the novel sources.  The 
local invariant solutions of the functional equations are constructed
in order to represent the counterterms for the one-loop subtractions.
The bleaching 
technique is fully extended to the fermion sector.
\par
The neutral sector of the vector mesons is analyzed in detail
in order to identify the physical fields for the photon and the
$Z$ boson. The identities necessary for the decoupling of the
unphysical modes are fully analyzed. 
These latter results are crucially
bound to the Landau gauge used throughout the paper.
\end{abstract}

\newpage
\section{Introduction}

In Ref.~\cite{Bettinelli:2008ey} 
  a consistent formulation of the electroweak  model based on a 
  nonlinear realization of the $SU(2)\otimes U(1)$ gauge group has been  
  presented by giving the tools required for the computation
 of radiative corrections in the loop expansion
(Feynman rules and the functional identities necessary in order to 
guarantee physical unitarity and to
carry out the subtraction procedure while respecting
the locality of the counterterms to every loop order).

  In the nonlinear realization there is no Higgs field \cite{higgs} in the
  perturbative spectrum.

In the present paper we consider the tree-level 
vertex functional by including all the 
required external
sources necessary for the functional formulation of the relevant
symmetries of the theory and for the symmetric subtraction
of the divergences.

The Local Functional Equation (LFE) 
\cite{Ferrari:2005ii,Bettinelli:2007tq}, which fixes 
  the 1-PI amplitudes involving at least one Goldstone boson 
  (descendant amplitudes) in terms of Goldstone-independent 1-PI amplitudes 
  (ancestor amplitudes), provides a hierarchy among
  1-PI Green functions. Once the ancestor amplitudes have been
  subtracted, the LFE uniquely fixes the descendant amplitudes.
The LFE holds together with the Slavnov-Taylor (ST) identity,
  which guarantees the fulfillment of physical unitarity 
  \cite{Ferrari:2004pd}, and the Landau Gauge Equation (LGE),
  which encodes the stability of the Landau gauge-fixing
  under radiative corrections.

In the present paper
the hierarchy is obtained by using the LFE and a  set of external
sources that ought to be complete in order to obtain all the descendant
amplitudes. The Weak Power-Counting (WPC) is derived
in the Landau gauge
 and used in the presence of this complete set
of sources. 
The peculiar behavior of the fermion UV dimension 
in the nonlinearly realized theory
is analyzed.
The method of bleaching is used for all fields and sources
with the aim of obtaining  the most general local solution of the
functional equations (STI, LFE and LGE). Finally
the construction of the complete effective action is performed
with the use of the WPC.
The subtraction procedure is then reconsidered
in the presence of the 
whole set of sources. The Ward-Takahashi identity (WTI)
associated to the electric charge is discussed in its consequences,
as the self-energy of the $\gamma-Z$ system and the description of
the photon field in physical amplitudes. 
The identities necessary for the decoupling of the
unphysical modes  in the Landau gauge are fully analyzed.

We find that the requirement of the validity of the WPC 
 imposes strong constraints
 on  the classical action of the nonlinearly realized 
 electroweak Standard Model.
 In fact  all possible symmetric anomalous couplings are
 forbidden by the WPC. Moreover two independent mass invariants 
 appear in the vector meson sector (thus relaxing the tree-level
 Weinberg relation between the masses of the $Z$ and $W$).

The symmetric finite  subtractions which are mathematically
allowed at higher orders in the loop expansion cannot
be reinserted back into the tree-level vertex functional
without violating either the symmetries or the WPC.
Therefore their interpretation as physical parameters
is {not possible} \cite{Bettinelli:2007zn}. 
One possible {\it Ansatz} is to perform Minimal Subtraction of properly normalized
1~-~PI amplitudes \cite{Bettinelli:2008ey,Bettinelli:2007tq}. 
We finally prove that 
this {\em Ansatz} guarantees the fulfillment of
all the relevant functional identities, order by order in the loop expansion.

The proof is based on a double grading expansion of the 1-PI amplitudes
in the number of loops and in the loop order of the counterterms.

The paper is organized as follows.
In Sect.~\ref{sec.classical} we introduce our notation
and provide a systematic construction of $SU(2)_L$-invariant
variables (bleaching procedure) in one-to-one
correspondence with the original gauge and matter fields.
The Feynman rules 
for the nonlinearly realized electroweak model are given
in Sect.~\ref{sec1.1}. 

In Sect.~\ref{sec.1.3} the gauge-fixing is performed in the Landau gauge. 
The BRST symmetry of 
the nonlinearly realized
theory is presented and the STI is obtained
by introducing the necessary anti-field external sources.
The LGE and the associated
ghost equation are also derived.
In Sect.~\ref{sec.lfe} the LFE is obtained
as a consequence of the invariance of the 
path-integral Haar measure
under local $SU(2)_L$ transformations.
The sources required in order to define at the renormalized level 
 the operators necessary for the LFE are also introduced. 
In Sect.~\ref{sec1.2} we show that the symmetry content
of the model allows for additional (anomalous) tree-level couplings. 

The WPC is discussed in Sect.~\ref{sec.wpc}.
In Sect.~\ref{sec.funct_id} we study the algebraic properties
of the linearized ST operator 
${\cal S}_0$ and of the linearized LFE operator ${\cal W}_0$. 
In Sect.~\ref{sec.bleach} the bleaching
procedure is extended to generate ${\cal S}_0$-invariant
variables.
These are relevant for the algebraic classification of the counterterms
order by order in the loop expansion.
Moreover we discuss the subtraction
procedure and the symmetric normalization of the 1-PI amplitudes.
In  Sect.~\ref{sec:neu} we consider the neutral sector of the
vector boson. A detailed study of the STI and of the LGE allows 
the identification of the physical fields of the photon and of
the $Z$ boson. Useful identities are derived in order
to verify the decoupling of the unphysical modes.
Finally conclusions are given in Sect.~\ref{sec.concl}.

Appendix~\ref{app.prop} collects the propagators in the Landau
gauge, while Appendix~\ref{app:C} is devoted to the technical
proof of the WPC.
Appendix ~\ref{app:neutral} contains the details
of the study of the neutral sector of the vector bosons.

\section{Classical symmetries and bleached variables}
\label{sec.classical}

The field content of the electroweak model based on
the nonlinearly realized $SU(2)_L\otimes U(1)$ gauge group
includes (leaving aside for the moment the ghosts
and the Nakanishi-Lautrup fields) the $SU(2)_L$
connection $A_\mu = A_{a\mu} \frac{\tau_a}{2}$
($\tau_a,~ a=1,2,3$ are the Pauli matrices), the 
$U(1)$ connection $B_\mu$, the fermionic left
doublets collectively denoted by $L$ and the right singlets, i.e.
\begin{eqnarray}
& L \in\Biggl\{ \left(
\begin{array}{r} l^u_{Lj}\\
l^d_{Lj}
\end{array} \right), \left(
\begin{array}{r}q^u_{Lj}\\
V_{jk}q^d_{Lk}
\end{array} \right), \quad j,k=1,2,3\Biggr\}, & \nonumber \\
& R \in\Biggl\{ \left(
\begin{array}{r}l^u_{Rj} \\
l^d_{Rj}
\end{array} \right), \left(
\begin{array}{r} q^u_{Rj}\\
q^d_{Rj}
\end{array} \right), \quad j = 1,2,3\Biggr\}. &
\label{sec.2.1}
\end{eqnarray} 
In the above equation the quark fields
$(q^u_j, j=1,2,3) = (u,c,t)$ and
$(q^d_j, j=1,2,3) = (d,s,b)$ are taken
to be the mass eigenstates in the tree-level
lagrangian; $V_{jk}$ is the CKM matrix.
Similarly we use for the leptons the notation
$(l^u_j, j=1,2,3) = (\nu_e,\nu_\mu,\nu_\tau)$ and
$(l^d_j, j=1,2,3) = (e,\mu,\tau)$.
The single left doublets are denoted by $L^l_j$,
$j=1,2,3$ for the leptons, $L^q_j$, $j=1,2,3$ for the 
quarks.
Color indexes are not displayed.

One also introduces the $SU(2)$ matrix $\Omega$
\begin{eqnarray}
\Omega = \frac{1}{v} (\phi_0 + i \phi_a \tau_a) \, , ~~~
\Omega^\dagger \Omega = 1 \Rightarrow \phi_0^2 + \phi_a^2 = v^2 \, .
\label{sec.2.2}
\end{eqnarray}
The  mass scale $v$  gives $\phi$ the canonical dimension
at $D=4$. We fix the direction of 
Spontaneous Symmetry Breaking
by imposing the tree-level constraint
\begin{eqnarray}
\phi_0 = \sqrt{v^2 - \phi_a^2} \, . 
\label{sec.2.3}
\end{eqnarray}
The condition $\langle \Omega \rangle = 1$ cannot be imposed
at a generic order of perturbation theory.

\par
The $SU(2)$ flat connection is defined by
\begin{eqnarray}
F_\mu = i \Omega \partial_\mu \Omega^\dagger \, .
\label{sec.2.4}
\end{eqnarray}
The transformation properties under the local $SU(2)_L$ transformations are
($g$ is the $SU(2)_L$ coupling constant)
\begin{eqnarray}
\begin{array}{ll}
 \Omega' = U \Omega \, , & B'_\mu = B_\mu \, , \\
 A'_\mu = U A_\mu U^\dagger + \frac{i}{g} U \partial_\mu U^\dagger \, ,
 & L' = U L \, , \\
 F'_\mu = U F_\mu U^\dagger + i U \partial_\mu U^\dagger \, , 
 & R'=R \, .
\end{array}
\label{sec.2.5}
\end{eqnarray}
Under local $U(1)_R$ transformations one has
\begin{eqnarray}
\begin{array}{ll}
 \Omega' = \Omega V^\dagger \, , & B'_\mu = B_\mu + \frac{1}{g'} \partial_\mu \alpha \, , \\
 A'_\mu = A_\mu ,
 & L' = \exp ( i \frac{\alpha}{2} Y_L) L \, , \\
 F'_\mu = F_\mu + i \Omega V^\dagger \partial_\mu V \Omega\, , 
 & R'=\exp ( i \frac{\alpha}{2} (Y_L + \tau_3) ) R  \, .
\end{array}
\label{sec.2.6}
\end{eqnarray}
where $V(\alpha) = \exp(i \alpha \frac{\tau_3}{2})$.

The electric charge is defined according to the Gell-Mann-Nishijima relation
\begin{eqnarray}
Q = I_3 + Y \, ,
\label{sec.2.6.bis}
\end{eqnarray}
where the hypercharge operator $Y$ is the generator of the
$U(1)_R$ 
transformations (\ref{sec.2.6})
and $I_3$ is an abstract object.
The introduction of the matrix $\Omega$ allows to perform an invertible
change of variables from the original set of fields to a new set
of $SU(2)_L$-invariant ones (bleaching procedure). 
For that purpose we define
\begin{eqnarray}
&& w_\mu = w_{a\mu} \frac{\tau_a}{2} = g \Omega^\dagger A_\mu \Omega - g' B_\mu \frac{\tau_3}{2}
+ i \Omega^\dagger \partial_\mu \Omega \, , \nonumber \\
&& \tilde L = \Omega^\dagger L \, .
\label{sec.2.7}
\end{eqnarray}
Both $w_\mu$ and $\tilde L$ are $SU(2)_L$-invariant, while under
$U(1)_R$ they transform as
\begin{eqnarray}
w'_\mu = V w_\mu V^\dagger \, , ~~~~
\tilde L' = \exp (i \frac{\alpha}{2} (\tau_3 + Y_L) ) \tilde L \, .
\label{sec.2.8}
\end{eqnarray}
I.e. the electric charge coincides with the hypercharge on the
bleached fields, as it is apparent from the comparison of eqs.(\ref{sec.2.6}),
(\ref{sec.2.6.bis}) and (\ref{sec.2.8}).

\subsection{Classical Action}\label{sec1.1}

Two mass invariants are expected for
the vector mesons, as a consequence of the breaking of the
global $SU(2)_R$ invariance induced by the hypercharge.
We introduce the charged combinations
\begin{eqnarray}
w^\pm_\mu = \frac{1}{\sqrt{2}} (w_{1\mu} \mp i w_{2\mu}) \, , ~~~ 
w^{\pm'}_\mu = \exp ( \pm i \alpha ) w^\pm_\mu \, .
\label{sec.2.9}
\end{eqnarray}
The neutral component $w_{3\mu}$ is invariant. Thus one obtains
two independent mass terms which can be parameterized as
\begin{eqnarray}
M^2 \Big ( w^+ w^- + \frac{1}{2} w_3^2 \Big ) \, , ~~~~ \frac{M^2 \kappa}{2} w_3^2 \, .
\label{sec.2.10}
\end{eqnarray}
Discarding the neutrino mass terms, the 
classical action for the nonlinearly realized $SU(2) \otimes U(1)$ gauge group with two 
independent mass parameters
for the vector mesons can be written as follows,
where the dependence on $\Omega$ is explicitly shown:
\begin{eqnarray}&&
S =
\Lambda^{(D-4)} \int d^Dx\,\Biggl( \,2 \,Tr\, \biggl\{
- \frac{1}{4}  G_{\mu\nu} G^{\mu\nu} - \frac{1}{4}  F_{\mu\nu}
F^{\mu\nu}
\Biggr\}
\nonumber\\&&
+M^2  \,Tr\, \biggl\{\bigl (gA_{\mu}
- \frac{g'}{2} \Omega\tau_3 B_\mu \Omega^\dagger
- F_{\mu}\bigr)^2\biggr\}
\nonumber\\&& 
+M^2\frac{ \kappa }{2}\Bigl( Tr\bigl\{(g \Omega^\dagger A_\mu \Omega 
- g' B_\mu \frac{\tau_3}{2}
+ i \Omega^\dagger \partial_\mu \Omega) \tau_3\bigr\}\Bigr)^2
\nonumber\\&&
+\sum_L\biggl[
\bar L \bigr(i\not\!\partial +g\not\!\!A 
+\frac{g'}{2}Y_L\not\!B\bigl)L\biggr]
+\sum_R\biggr[\bar R \bigr(i\not\!\partial 
+ \frac{g'}{2} (Y_L+\tau_3) \not\!B\bigl)R
\biggr]
\nonumber \\&&
+\sum_j\biggl[
m_{l_j}~\bar R^l_j\frac{1-\tau_3}{2}\Omega^\dagger L^l_j
-
m_{q^u_j}~\bar R^q_j \frac{1+\tau_3}{2}\Omega^\dagger L^q_j
\nonumber\\&&
+
m_{q^d_k} V^\dagger_{kj} ~\bar R^q_k
\frac{1-\tau_3}{2}\Omega^\dagger 
 L^q_j +h.c.
\biggr]
\Biggr) \, .
\label{pre.1}
\end{eqnarray}
In $D$ dimensions the doublets $L$ and $R$ obey
\begin{eqnarray}
\gamma_{_D} L = - L \quad
\gamma_{_D} R =  R,
\label{pre.2}
\end{eqnarray}
being $\gamma_{_D}$ a gamma matrix that anti-commutes with every other $\gamma^\mu$.

The non-Abelian field strength $G_{\mu\nu}$ is defined by
\begin{eqnarray}
G_{\mu\nu}= G_{a\mu\nu} \frac{\tau_a}{2} = 
(\partial_\mu A_{a\nu} - \partial_\nu A_{a\mu}
+ g \epsilon_{abc} A_{b\mu} A_{c\nu}) \frac{\tau_a}{2} \, ,
\label{n.ab.fs.}
\end{eqnarray}
while the Abelian field strength $F_{\mu\nu}$ is
\begin{eqnarray}
F_{\mu\nu} = \partial_\mu B_\nu - \partial_\nu B_\mu \, .
\label{ab.fs}
\end{eqnarray}
In the above equation the phenomenologically
successful  structure of the couplings
has been imposed by hand. The discussion of the
possible anomalous couplings and of the
stabilization mechanism induced by the 
WPC is deferred to Sect.~\ref{sec1.2}.

\subsection{Gauge-fixing and BRST symmetry}\label{sec.1.3}

In order to set up the framework for the perturbative
quantization of the model,
the classical action in eq.(\ref{pre.1}) needs to be gauge-fixed.
The ghosts associated with the $SU(2)_L$ symmetry are denoted
by $c_a$. 
Their anti-ghosts are denoted by $\bar c_a$, the Nakanishi-Lautrup
fields by $b_a$.
It is also useful to adopt the matrix notation
\begin{eqnarray}
c = c_a \frac{\tau_a}{2} \, , ~~~~ b = b_a \frac{\tau_a}{2} \, , ~~~~
\bar c =  \bar c_a \frac{\tau_a}{2} \, .
\label{sec.3.1}
\end{eqnarray}
The abelian ghost  is $c_0$, the abelian anti-ghost $\bar c_0$
and the abelian Nakanishi-Lautrup field $b_0$.

For the sake of simplicity we deal here with the Landau gauge.
We also include the anti-fields for
the $SU(2)_L$ BRST transformation (those for the $U(1)_R$ BRST transformation
are not required since the Abelian ghost is free in the Landau gauge).
\begin{eqnarray}&&
\Gamma^{(0)}_{\rm GF}
\nonumber\\&&
=
\Lambda^{(D-4)}\int d^Dx \Biggl( b_0 \partial_\mu B^\mu 
-\bar c_0 \Box c_0 + 2 Tr~\Bigl\{  b\partial_\mu A^\mu 
-\bar c\partial^\mu D[A]_\mu c
\nonumber\\&&
+ V^\mu ~\biggl ( D[A]_\mu b 
-ig \bar cD[A]_\mu c - ig (D[A]_\mu c) \bar c
\biggr)
+\Theta^\mu~D[A]_\mu\bar c\Bigr\} +K_0\phi_0 
\nonumber\\&& 
+ A^*_{a\mu} \mathfrak{s}A^\mu_a + 
\phi_0^* \mathfrak{s} \phi_0 + 
\phi_a^* \mathfrak{s} \phi_a + c_a^* \mathfrak{s} c_a
+ \sum_L\Big ( L^* \mathfrak{s}L + 
\bar L^* \mathfrak{s}\bar{ L}\Bigr)
\Biggr) 
\label{pre.19}
\end{eqnarray}
The full tree-level vertex functional is 
\begin{eqnarray}
\G^{(0)} = S + \G^{(0)}_{GF} \, .
\label{new.1}
\end{eqnarray}
The $SU(2)_L$ BRST symmetry is generated by the differential
$\mathfrak{s}$:
\begin{equation}
\begin{array}{llll}
\mathfrak{s} A_\mu = D[A]_\mu ~ c
&
\mathfrak{s} \Omega = ig~ c ~ \Omega~~~
&
\mathfrak{s} \bar c = b 
&
\mathfrak{s} \bar c_0 = 0 
\\
\mathfrak{s} c = i g~ c ~ c
&
\mathfrak{s} B_\mu = 0
&
\mathfrak{s} b = 0
&
\mathfrak{s} b_0 = 0
\\
\mathfrak{s}L = igcL
&
\mathfrak{s}R =0
&
\mathfrak{s}c_0=0 \, .
&
\end{array} 
\label{pre.20}
\end{equation}
The  source $K_0$ is required in order
to define the nonlinear constraint $\phi_0$.
This implies the inclusion of the source $\phi_0^*$, coupled to 
the BRST variation of $\phi_0$.
The resulting STI is
\begin{eqnarray}
&& \!\!\!\!\!\!\!\!\!\!
{\cal S}\G \equiv \int d^Dx \,\Biggl[ \Lambda^{-(D-4)}\Big (
 \G_{ A^*_{a\mu}}  \G_{ A_a^\mu}
+
 \G_{ \phi_a^*}  \G_{ \phi_a}
+ 
 \G_{ c_a^*} \G_{ c_a}
 \nonumber \\
&&
+ \G_{ L^*}  \G_{ L}
+ \G_{ \bar L^*} 
 \G_{ \bar L}\Big )
+ b_a  \G_{ \bar c_a}
 + \Theta_{a\mu}  \G_{ V_{a\mu}}
      - K_0  \G_{ \phi_0^*} 
\Biggr] = 0 \, .
\label{brst.13}
\end{eqnarray}
In the above equation the background connection $V_{a\mu}$ is paired into 
a doublet with $\Theta_{a\mu}$. This is a
standard procedure in order to guarantee the independence of the physics on the
background sources \cite{Ferrari:2000yp}.
$(\phi_0^*, -K_0)$  are also arranged
into doublets in the above STI. 
This is required in order to preserve the STI in the
presence of the source $K_0$ and signals that $K_0$
is not a physical variable. 
This feature has been 
addressed in \cite{Bettinelli:2007tq} in the context of massive
$SU(2)$ Yang-Mills theory.

Moreover the following Abelian STI holds:
\begin{eqnarray}
&&\!\!\!\!\!\!\!\!\!\!
- {\frac{2}{g'}}\Lambda^{(D-4)} \Box b_0
- {\frac{2}{g'}}\partial^\mu  \frac{\delta\G }{\delta  B^\mu}
- {\Lambda^{(D-4)}}\phi_3 K_0 
+ \phi_2 \frac{\delta\G }{\delta \phi_1}
-\phi_1\frac{\delta\G }{\delta \phi_2}
-{\frac{1}{\Lambda^{(D-4)}}}
\frac{\delta\G }{\delta K_0}\frac{\delta\G }{\delta \phi_3}
\nonumber\\&&
-\phi_3^*\frac{\delta\G }{\delta \phi_0^*}
+ \phi_2^* \frac{\delta\G }{\delta \phi_1^*}
-\phi_1^* \frac{\delta\G }{\delta \phi_2^*}
+\phi_0^*\frac{\delta\G }{\delta \phi_3^*} 
\nonumber\\&&
+iY_LL\frac{\delta\G }{\delta L}
-iY_L\bar L\frac{\delta\G }{\delta \bar L}
+i(Y_L+\tau_3)R\frac{\delta\G }{\delta R}
-i\bar R(Y_L+\tau_3)\frac{\delta\G }{\delta \bar R}
\nonumber\\&&
-iY_LL^*\frac{\delta\G }{\delta L^*}
+iY_L\bar L^*\frac{\delta\G }{\delta \bar L^*}
=0.
\label{st.3}
\end{eqnarray}
The transformations of the fields in the above equation
are generated by the $U(1)_R$ BRST symmetry
\begin{equation}
\begin{array}{llll}
\mathfrak{s}_1 A_\mu = 0
&
\mathfrak{s}_1 \Omega =- \frac{i}{2} g' \Omega c_0 \tau_3
&
\mathfrak{s}_1 \bar c=0
&
\mathfrak{s}_1 \bar c_0 = b_0
\\

\mathfrak{s}_1 c=0
&
\mathfrak{s}_1 B_\mu = \partial_\mu  c_0
&
\mathfrak{s}_1 b=0
&
\mathfrak{s}_1 b_0 = 0.
\\
\mathfrak{s}_1 L= \frac{i}{2} g'c_0Y_LL
&
\mathfrak{s}_1 R = \frac{i}{2} g'c_0(Y_L+\tau_3) R
&
\mathfrak{s}_1 c_0 = 0
&
\end{array} 
\label{pre.21}
\end{equation}
By construction 
\begin{eqnarray}
\{ \mathfrak{s}, \mathfrak{s}_1 \} = 0 \, .
\label{comp.1}
\end{eqnarray}
Eq.(\ref{st.3}) can be derived from the
invariance under the $U(1)_R$ transformations
in eq.(\ref{sec.2.6}) supplemented by the following transformations
on the additional variables (we set
$\Omega^* = \phi_0^* - i \phi_a^* \tau_a$)
\begin{equation}
\begin{array}{llll}
V_\mu' = V_\mu 
&
\Omega^{*'} = {V\Omega^* }
&
L^{*'} = \exp(-i\frac{\alpha}{2} Y_L) L^*
&
K_0' = K_0
\\
\Theta_\mu'=\Theta_\mu
&
b'=b
&
\bar L^{*'} = \exp(i\frac{\alpha}{2} Y_L) \bar L^*
&
 b_0' = b_0
\\
c'=c
&
\bar c' = \bar c
&
c^{*'}=c^*
&
\\ 
c_0' = c_0
&
\bar c_0' = \bar c_0
&
A_\mu^{*'}= A_\mu^{*}.
&
\end{array} 
\label{st.3.1}
\end{equation}

{The ghost number is defined as follows:
$A^*_{a\mu}, \phi_a^*,\phi_0^*,
L^*, {\bar L}^*, \bar c_a, \bar c_0$ have ghost number -1,
$c^*$ has ghost number -2, 
$c_a$, $c_0$ and $\Theta_{a\mu}$ have
ghost number +1, while all the other
fields and external sources have ghost number zero.
}

The LGE  is
\begin{eqnarray}
\G_{ b_a} =  \Lambda^{(D-4)} \Big ( 
 D^\mu[V](A_\mu - V_\mu)\Big )_a 
\label{b.eq}
\end{eqnarray}
which implies the ghost equation
\begin{eqnarray}
 \G_{ \bar c_a} =   
\Big ( -D_\mu[V]  \G_{ A_{\mu}^*} 
+ \Lambda^{(D-4)}  D_\mu[A] \Theta^\mu
\Big )_a \, ,
\label{gh.eq}
\end{eqnarray}
by using the STI (\ref{brst.13}).

\subsection{The Local Functional Equation}\label{sec.lfe}

The dependence of the vertex functional on the
Goldstone fields is controlled by the LFE 
associated to the invariance of the path-integral
Haar measure under the $SU(2)_L$ transformations
in eq.(\ref{sec.2.5}), extended to the ghost, anti-ghost,
Nakanishi-Lautrup fields and to the external sources 
according to
\begin{equation}
\begin{array}{llll}
V'_{\mu}= UV_{\mu}U^\dagger + \frac{i}{g} U\partial_\mu U^\dagger
&
\Omega^{*'} = {\Omega^*U^\dagger }
&
L^{*'} =  L^*U^\dagger
&
K_0' = K_0
\\
\Theta_\mu'=U\Theta_\mu U^\dagger
&
b'=UbU^\dagger
&
\bar L^{*'} = U \bar L^*
&
 b_0' = b_0
\\
c'=Uc U^\dagger
&
\bar c' = U \bar c U^\dagger
&
c^{*'}=c^*
&
\\ 
c_0' = c_0
&
\bar c_0' = \bar c_0
&
A_\mu^{*'}= UA_\mu^{*} U^\dagger.
&
\end{array} 
\label{pre.6.1}
\end{equation}
Thus the resulting identity associated to the $SU(2)_L$
local transformations is ($x$-dependence is not shown)
\begin{eqnarray}
&&({\cal W}\G)_{a} \equiv 
-{\frac{1}{g}}\partial_\mu \G_{ V_{a \mu}} 
+ \epsilon_{abc} V_{c\mu} \G_{ V_{b\mu}}
-{\frac{1}{g}}\partial_\mu \G_{ A_{a \mu}}
\nonumber \\&&  
+ \epsilon_{abc} A_{c\mu} \G_{ A_{b\mu}}
+ \epsilon_{abc} b_c \G_{ b_b}
+ \frac{{\Lambda^{(D-4)}}}{2} K_0\phi_a
+ \frac{1}{2\Lambda^{(D-4)}} \G_{ K_0} 
\G_{ \phi_a} 
\nonumber \\
&&  
+  
\frac{1}{2} \epsilon_{abc} \phi_c \G_{ \phi_b} 
      + \epsilon_{abc} \bar c_c \G_{ \bar c_b}
      + \epsilon_{abc} c_c \G_{  c_b}
 \nonumber \\&&
{
+\frac{i}{2}\tau_aL\G_{  L}
-\frac{i}{2}\bar L\tau_a\G_{ \bar  L}
-\frac{i}{2}L^*\tau_a\G_{  L^*}
+\frac{i}{2}\tau_a\bar L^*\G_{ \bar  L^*}
}
 \nonumber \\
&& + \epsilon_{abc} \Theta_{c\mu} \G_{ \Theta_{b\mu} }
      + \epsilon_{abc} A^*_{c\mu} \G_{ A^*_{b\mu}}
      + \epsilon_{abc} c^*_c \G_{  c^*_b} 
 - \frac{1}{2} \phi_0^* \G_{ \phi^*_a}
\nonumber \\
&& +  
\frac{1}{2} \epsilon_{abc} \phi^*_c \G_{ \phi^*_b} 
+ \frac{1}{2} \phi_a^* \G_{ \phi_0^*}
= 0 \, ,
\label{bkgwi}
\end{eqnarray}
where the nonlinearity of the realization of the $SU(2)_L$ 
gauge group is revealed by the presence of the bilinear term
{$ \G_{ K_0} \G_{ \phi_a} $}.
Since in the loop-wise expansion $\G_{K_0}$ is invertible,
eq. (\ref{bkgwi}) entails 
that every amplitude with $\phi-$external  leg 
(descendant amplitudes) can be obtained from those without. 

This is a crucial property in order to tame the divergences
of the model. In fact already at one loop level the
Feynman rules in eq.(\ref{new.1}) give rise to
divergent Feynman diagrams with an arbitrary number
of external $\phi$-legs.
However at every loop order there is only a finite
number of ancestor amplitudes, i.e. amplitudes
which do not involve external Goldstone fields.
This property is referred to as the WPC.
Consequently a finite number of subtractions is required in order
to make the theory finite at each loop order.

\section{Anomalous Couplings}\label{sec1.2}

Any $U(1)_R$-invariant local functional
built out of the components of $w_\mu, \tilde L, R$, the
abelian field strength $F_{\mu\nu} = 
\partial_\mu B_\nu - \partial_\nu B_\mu$ and derivatives
thereof (covariant derivatives w.r.t. $B_\mu$ for $U(1)_R$-charged
fields, ordinary derivatives for the neutral fields) is allowed
on symmetry grounds.

We discuss here those invariants with dimension $\leq 4$.

Many possibilities arise for the interaction terms. For the gauge bosons
self-interactions 
\begin{eqnarray}
&& a_1 (w^+ w^-)^2 \, , ~~~~ a_2 (w^+)^2 (w^-)^2 \, , ~~~~ a_3 (w^+ w^-) w_3^2 \, ,
\nonumber \\
&& a_4 ~ w_{3\nu} \partial^\mu w^{+\nu} w^-_\mu \, , ~~~~
   a_5 ~ w_{3\nu} w^{+\nu} \partial w^- \, , \nonumber \\
&&   a_6 ~ w_3^\nu w^+_\mu \partial^\mu w^-_\nu \, , ~~~~~
   a_7 ~ w_{3\nu} \partial w^+ w^{-\nu} \, , \nonumber \\
&& a_8 ~ w_{3\nu} \partial^\nu w^+_\mu w^{-\mu} \, , ~~~~
   a_9 ~ w_{3\nu} w^+_\mu \partial^\nu w^{-\mu} \, .
\label{sec.2.12}
\end{eqnarray}
Hermiticity requires $a_4^*=a_6$, $a_5^*=a_7$ and  $a_8^*=a_9$.
For the leptonic neutral currents 
\begin{eqnarray}
&& g^{Lu,0}_{kj} ~ \bar l^u_{Lk} \slsh{w}_3 l^u_{Lj} \, , ~~~
g^{Ld,0}_{kj} ~ \bar l^d_{Lk} \slsh{w}_3 l^d_{Lj} \, , \nonumber \\
&& g^{Ru,0}_{kj} ~ \bar l^u_{Rk} \slsh{w}_3 l^u_{Rj} \, , ~~~
g^{Rd,0}_{kj} ~ \bar l^d_{Rk} \slsh{w}_3 l^d_{Rj} \, .
\label{sec.2.13}
\end{eqnarray}
A similar pattern applies to the quark neutral currents:
\begin{eqnarray}
&& h^{Lu,0}_{kj} ~ \bar q^u_{Lk} \slsh{w}_3 q^u_{Lj} \, , ~~~
h^{Ld,0}_{kj} ~ \bar q^d_{Lk} \slsh{w}_3 q^d_{Lj} \, , \nonumber \\
&& h^{Ru,0}_{kj} ~ \bar q^u_{Rk} \slsh{w}_3 q^u_{Rj} \, , ~~~
h^{Rd,0}_{kj} ~ \bar q^d_{Rk} \slsh{w}_3 q^d_{Rj} \, .
\label{sec.2.14}
\end{eqnarray}
For the charged currents one has in the leptonic sector
\begin{eqnarray}
&& g^{Lu,+}_{kj} ~ \bar l^u_{Lk} \slsh{w}^+ l^d_{Lj} + \mbox{h.c.} \, , ~~~
g^{Ru,+}_{kj} ~ \bar l^u_{Rk} \slsh{w}^+ l^d_{Rj}  + \mbox{h.c.} \, , ~~~
\label{sec.2.14.bis}
\end{eqnarray}
and in the hadronic sector
\begin{eqnarray}
&& h^{Lu,+}_{kj} ~ \bar q^u_{Lk} \slsh{w}^+ q^d_{Lj} + \mbox{h.c.} \, , ~~~
h^{Ru,+}_{kj} ~ \bar q^u_{Rk} \slsh{w}^+ q^d_{Rj} + \mbox{h.c.}
\label{sec.2.15}
\end{eqnarray}
The anomalous gauge bosons couplings in eq.(\ref{sec.2.12}) 
are not forbidden  on symmetry grounds, as well as
the flavor-changing neutral currents generated by the
off-diagonal elements of the couplings matrices in
eqs.(\ref{sec.2.13}) and (\ref{sec.2.14}).
They are excluded by hand in eq.(\ref{new.1})
on phenomenological grounds. In Sect.~\ref{sec.wpc}
we show that this choice is unique if one requires
the weak power-counting formula (\ref{wpc}).

\section{The Weak Power-Counting}\label{sec.wpc}

In the massive nonlinearly realized $SU(2)$
Yang-Mills theory \cite{Veltman:1968ki} 
and in the Electroweak model based on the
nonlinear representation of the 
$SU(2) \otimes U(1)$ gauge group 
\cite{Bettinelli:2008ey}
the number of divergent 1-PI amplitudes involving the Goldstone fields 
is infinite already at one loop. 
However these amplitudes
are uniquely fixed order by order in the loop
expansion by the LFE in eq.(\ref{bkgwi}) once the 1-PI
amplitudes not involving the Goldstone fields (ancestor amplitudes)
are known. We call this property hierarchy among 1-PI Green functions. 
It holds in the nonlinear sigma model 
in the flat connection formalism \cite{Ferrari:2005ii}. 
The tools for the integration of the LFE have been developed in
\cite{Bettinelli:2007kc}.
Hierarchy among 1-PI Green functions  has been studied 
for the massive nonlinearly realized SU(2) Yang-Mills theory 
in \cite{Bettinelli:2007tq}.

The WPC \cite{Ferrari:2005va} amounts to the request that 
only a finite number of 
divergent ancestor amplitudes exists at each loop order.
This restricts the number of allowed tree-level interaction
vertexes.

Let ${\cal G}$ 
be an arbitrary $n$-loop 1-PI ancestor 
graph with $I$ internal lines, $V$ vertexes and a given set 
$\{ N_A, N_B, N_F, N_{\bar F}, N_c, N_V, N_\Theta, N_{\phi_0^*},
N_{K_0}, N_{\phi_a^*}, N_{A^*}, N_{c^*}, N_{L^*}, N_{\bar L^*} \}$ of external legs.
\noindent
$F,\bar F$ are a collective notation for the fermion and anti-fermion
matter fields, which can be treated in a unified manner.
Then the superficial degree of divergence of the graph ${\cal G}$
is bounded by 
\begin{eqnarray}
&& d({\cal G}) \leq (D-2)n +2 - N_A - N_B - N_c - N_F 
- N_{\bar F}  -  N_V - N_{\phi_a^*} \nonumber \\
&& ~~~~ - 2 (N_\Theta + N_{A^*} + N_{\phi_0^*} + N_{L^*} + N_{{\bar L}^*} + N_{c^*} + N_{K_0} ) \, .
\label{wpc}
\end{eqnarray}
The proof of this formula is given in Appendix~\ref{app:C} by exploiting the symmetric formalism where 
the original fields $(A_{a\mu}, B_\mu)$ are used instead of the mass eigenstates $W^\pm_\mu, Z_\mu, A_\mu$.
The propagators in the symmetric formulation are summarized in Appendix~\ref{app.prop}.

\begin{figure}
\begin{center}
\includegraphics[width=4.5 truecm]{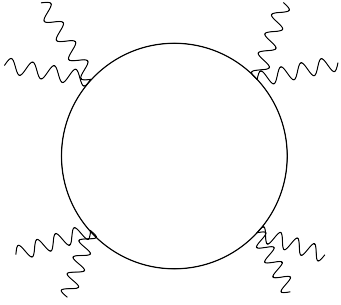}
\end{center}
\caption{Logarithmically divergent one-loop graphs with an arbitrary number of external gauge boson legs 
(solid lines denote Goldstone propagators)}
\label{fig.0}
\end{figure}

The validity of the WPC formula forbids the appearance of the
anomalous gauge bosons self-interactions in eq.(\ref{sec.2.12})
into the tree-level vertex functional $\G^{(0)}$ 
in eq.(\ref{new.1}).
In fact the terms in eq.(\ref{sec.2.12}) would give rise 
upon expansion in powers of the Goldstone fields to
quadrilinear interaction vertexes with two gauge bosons,
two Goldstone legs and two derivatives. Therefore
at one loop level there would exist an infinite number
of divergent amplitudes with external gauge boson legs,
associated to graphs like the one in Figure ~\ref{fig.0}.
Therefore the WPC would be maximally violated already
at one loop level.

The only allowed combination is the Yang-Mills action,
as was pointed out in \cite{Bettinelli:2007tq}.
On the other hand, the WPC does not put any constraint on the 
gauge boson mass invariants.
In the nonlinearly realized electroweak model the
hypercharge $U(1)_R$ invariance 
allows for the two independent mass terms in eq.(\ref{sec.2.10}).

\begin{figure}
\begin{center}
\includegraphics[width=5truecm]{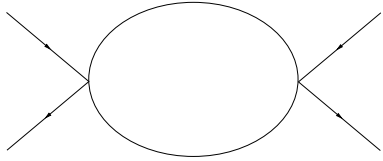}
\end{center}
\caption{Logarithmically divergent one-loop graphs with four fermion legs
(solid lines denote Goldstone propagators)}
\label{fig.1}
\end{figure}

According to the WPC formula in eq.(\ref{wpc})  the fermionic fields
have UV degree $1$ (instead of $3/2$ as in  power-counting 
renormalizable theories).
This is a peculiar feature of the electroweak model based on the nonlinearly realized
gauge group $SU(2) \otimes U(1)$.
It is easy to see that the UV degree of massive chiral fermions 
in the nonlinearly realized theory cannot be greater than $1$.
In fact the invariant fermionic mass terms in
eq.(\ref{pre.1})  contain couplings generated by the expansion of
the nonlinear constraint $\phi_0$ with the following structure
\begin{eqnarray}
\frac{m_f}{v} \bar f f \phi_0 \sim m_f \bar f f \Big [1 -  \sum_{k=1}^\infty 
\frac{1}{k!}   \frac{(2k-3)!!}{2^k} \Big ( \frac{ \phi^2}{v^2} \Big )^k  \Big ] \, .
\label{coup.1}
\end{eqnarray}
The first interaction term on the R.H.S. contains a quadrilinear coupling 
giving rise to graphs like the one in Figure~\ref{fig.1}. Thus there are 
one loop logarithmically divergent 
graphs with four external fermion legs and therefore 
the UV degree of massive chiral fermions can be at most one.
For massless neutrinos the bond of eq.(\ref{wpc}) 
still works but one cannot associate their UV dimension
on the basis of the degree of divergence of the graphs
in Figure~\ref{fig.1}.

\begin{figure}
\begin{center}
\includegraphics[width=6.5truecm]{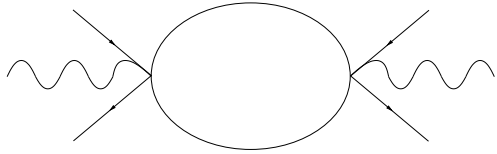}
\end{center}
\caption{Logarithmically divergent one-loop graphs with four fermionic external legs 
and two gauge bosons legs generated by $\bar l^u_{Lk} \slsh{w}_3 l^u_{Lj}$
(solid lines denote Goldstone propagators)}
\label{fig.2}
\end{figure}

If the symmetric interactions in eqs.(\ref{sec.2.13})-(\ref{sec.2.15}) 
are turned on,
the UV degree of the fermions is downgraded to one half.
This is readily established by expanding the invariants in powers of the
Goldstone fields and by looking at the graphs arising from the interaction
vertexes involving two Goldstone legs.
An example is displayed in Figure ~\ref{fig.2}.

\begin{figure}
\begin{center}
\includegraphics[width=5truecm]{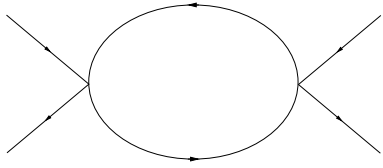}
\end{center}
\caption{Quadratically divergent one-loop graphs with four fermionic external legs 
generated by four-fermion interactions}
\label{fig.3}
\end{figure}

It is interesting to notice that fermions with UV degree equal to one half 
are compatible with four fermion interactions generated in a symmetric way 
by using invariant bleached variables,  like for instance
\begin{eqnarray}
\bar l^u_{Rj} \tilde l^u_{Lj} \bar l^u_{Rj} \tilde l^u_{Lj} + \mbox{h.c.}
\label{coup.2}
\end{eqnarray}
which would generate the 
quadratically divergent one loop graph in Figure~\ref{fig.3}.

\medskip
In the nonlinearly realized theory it turns out that
 one is the  UV degree for the fermion fields 
compatible with the invariant mass terms for chiral fermions.
As a consequence one recovers via the WPC
the phenomenologically successful structure of the SM couplings
in eq.(\ref{pre.1}).

\section{Functional Identities and Minimal Subtraction Procedure}\label{sec.funct_id}

Perturbation theory is carried out in the loop-wise expansion.
Accordingly the functional identities in eq.(\ref{brst.13}), (\ref{st.3}),
(\ref{b.eq}), (\ref{gh.eq}) and (\ref{bkgwi}) are developed
order by order in $\hbar$. We denote by $\G^{(n)}$ the 
$n$-th loop vertex functional, i.e.
\begin{eqnarray}
\G = \sum_{n=0}^\infty \G^{(n)} \, .
\label{fid.1}
\end{eqnarray}
By eq.(\ref{b.eq}) $\G^{(n)}$, $n\geq 1$ is independent of $b_a$.
By eq.(\ref{gh.eq}) the dependence of $\G^{(n)}$, $n\geq 1$
on $\bar c_a$ only happens via the combination
\begin{eqnarray}
\hat A_{a\mu}^* = A^*_{a\mu} + (D_\mu[V]\bar c)_a \, .
\label{fid.2}
\end{eqnarray}
At order $n \geq 1$ in the loop expansion the STI 
in eq.(\ref{brst.13}) is
\begin{eqnarray}
{\cal S}_0 (\G^{(n)}) + \sum_{j=1}^{n-1} (\G^{(n-j)},\G^{(j)}) = 0 \, ,
\label{st.fid.1}
\end{eqnarray}
where the classical linearized ST operator ${\cal S}_0$ is
given by
\begin{eqnarray}
&& \!\!\!\!\!\!\!\!\!\!
{\cal S}_0\G \equiv \int d^Dx \,\Biggl[ \Lambda^{-(D-4)}\Big (
\G^{(0)}_{ A_a^\mu} 
\frac{\delta }{\delta A^*_{a\mu}}
+
\G^{(0)}_{ A^*_{a\mu}} \frac{\delta}{\delta A_a^\mu}
+
\G^{(0)}_{ \phi_a^*} \frac{\delta }{\delta \phi_a}
+
 \G^{(0)}_{ \phi_a}\frac{\delta }{\delta \phi_a^*}
 \nonumber \\
&&
+ 
\G^{(0)}_{ c_a^*}\frac{\delta }{\delta c_a}
+ 
\G^{(0)}_{ c_a}\frac{\delta }{\delta c_a^*}
+\G^{(0)}_{ L^*} \frac{\delta }{\delta L}
+ \G^{(0)}_{ L}\frac{\delta }{\delta L^*}
 \nonumber \\
&&
+\G^{(0)}_{ \bar L^*} 
\frac{\delta }{\delta \bar L}
+\G^{(0)}_{ \bar L}
\frac{\delta }{\delta \bar L^*} \Big )
+ b_a \frac{\delta }{\delta \bar c_a}
 + \Theta_{a\mu} \frac{\delta }{\delta V_{a\mu}}
      - K_0 \frac{\delta }{\delta \phi_0^*} 
\Biggr]\G \, .
\label{brst.14}
\end{eqnarray}
The bracket in eq.(\ref{st.fid.1}) is 
\begin{eqnarray}
(X,Y) = \int d^Dx \, \Lambda^{-(D-4)} \sum_j \frac{\delta X}{\delta \varphi_j^*} \frac{\delta Y}{\delta \varphi_j} \, ,
\label{bracket.1}
\end{eqnarray}
where $\varphi \in \{ A_{a\mu}, \phi_a, c_a, L, \bar L\}$ and
$\varphi^*_j$ stands for the anti-field associated to~$\varphi_j$.

At order $n\geq 1$ the LFE in eq.(\ref{bkgwi}) yields
\begin{eqnarray}
({\cal W}_0 \G^{(n)})_a + \frac{1}{2 \Lambda^{(D-4)}} \sum_{j=1}^{n-1} 
\frac{\delta \G^{(n-j)}}{\delta K_0(x)} \frac{\delta \G^{(j)}}{\delta \phi_a(x)} =  0
\label{bracket.2}
\end{eqnarray}
where ${\cal W}_0$ is the classical linearized version of ${\cal W}$:
\begin{eqnarray}
&&({\cal W}_0\G)_a \equiv \Biggl(
-{\frac{1}{g}}\partial_\mu \frac{\delta }{\delta V_{a \mu}} 
+ \epsilon_{abc} V_{c\mu} \frac{\delta }{\delta V_{b\mu}}
-{\frac{1}{g}}\partial_\mu \frac{\delta }{\delta A_{a \mu}}
\nonumber \\&&  
+ \epsilon_{abc} A_{c\mu} \frac{\delta }{\delta A_{b\mu}}
+ \epsilon_{abc} b_c \frac{\delta }{\delta b_b}
+ \frac{1}{2\Lambda^{(D-4)}}   \frac{\delta \G^{(0)}}{\delta K_0}
\frac{\delta }{\delta \phi_a}
\nonumber \\
&&  
+ \frac{1}{2\Lambda^{(D-4)}} 
\frac{\delta \G^{(0)}}{\delta \phi_a} 
\frac{\delta }{\delta K_0} 
+  
\frac{1}{2} \epsilon_{abc} \phi_c \frac{\delta }{\delta \phi_b} 
      + \epsilon_{abc} \bar c_c \frac{\delta }{\delta \bar c_b}
      + \epsilon_{abc} c_c \frac{\delta }{\delta  c_b}
 \nonumber \\&&
+\frac{i}{2}\tau_aL\frac{\delta}{\delta  L}
-\frac{i}{2}\bar L\tau_a\frac{\delta}{\delta \bar  L}
-\frac{i}{2}L^*\tau_a\frac{\delta}{\delta  L^*}
+\frac{i}{2}\tau_a\bar L^*\frac{\delta}{\delta \bar  L^*}
 \nonumber \\&& 
+ \epsilon_{abc} \Theta_{c\mu} \frac{\delta }{\delta \Theta_{b\mu} }
      + \epsilon_{abc} A^*_{c\mu} \frac{\delta }{\delta A^*_{b\mu}}
      + \epsilon_{abc} c^*_c \frac{\delta }{\delta  c^*_b} 
 - \frac{1}{2} \phi_0^* \frac{\delta }{\delta \phi^*_a}
\nonumber \\
&& +  
\frac{1}{2} \epsilon_{abc} \phi^*_c \frac{\delta }{\delta \phi^*_b} 
+ \frac{1}{2} \phi_a^* \frac{\delta }{\delta \phi_0^*} 
\Biggr)\G \, .
\label{bkgwi.1}
\end{eqnarray}
It is straightforward to prove that
\begin{eqnarray}
[{\cal S}_0,{\cal W}_0] = 0.
\label{fle.5}
\end{eqnarray}
%
\subsection{Bleached Variables}\label{sec.bleach}

The LFE in eq.(\ref{bracket.2}) can be explicitly 
integrated (with no locality restrictions) by using the
techniques developed in \cite{Bettinelli:2007kc}.

The first step is to extend the bleaching technique 
in order to generate variables  invariant  under ${\cal W}_0$.
This has been done for massive $SU(2)$ Yang-Mills theory
in \cite{Bettinelli:2007tq}. Here we provide the extension to the case
of chiral fermions.

Along the lines of \cite{Bettinelli:2007tq} we introduce
the bleached partners of $c$ and of the external sources:
\begin{eqnarray}
& v_\mu =  g \Omega^\dagger V_\mu \Omega - g' B_\mu \frac{\tau_3}{2}
+ i \Omega^\dagger \partial_\mu \Omega \, ,
& \tilde \Theta_\mu = \Omega^\dagger \Theta_\mu \Omega \, , \nonumber \\
& \tilde \Omega^* = \Omega^\dagger \Omega^* \, , 
& \tilde c = \Omega^\dagger c \Omega \, , \nonumber \\
& \widetilde{\hat A^*_\mu} = \Omega^\dagger \hat A^*_\mu
\Omega \, , & \tilde c^* = \Omega^\dagger c^* \Omega \, ,
\nonumber \\
& \tilde L^* = L^* \Omega \, , & \tilde {\bar L}^* = \Omega^\dagger 
\bar L^* \, .
\label{q.bleach.1}
\end{eqnarray}
The invariance of the above variables under ${\cal W}_0$
follows directly from eq.(\ref{pre.6.1}).
Moreover it can be proved \cite{Bettinelli:2007tq} that 
the following combination is ${\cal W}_0$-invariant:
\begin{eqnarray}
\widetilde K_0 = \frac{1}{v} \left ( \Lambda^{D-4} \frac{v^2  K_0}{\phi_0}
- \phi_a \frac{\delta}{\delta \phi_a} 
\left(\left . \G^{(0)} \right |_{K_0=0}\right ) \right ) \, .
\label{q.bleach.2}
\end{eqnarray}
The bleached variables in eq.(\ref{sec.2.7}) are ${\cal W}_0$-invariant.
The operator ${\cal W}_0$ takes a particularly simple form 
in the bleached variables:
\begin{eqnarray}
({\cal W}_0 \G)_a  = \Theta_{ab} \frac{\delta}{\delta \phi_b} \G \, ,
\label{q.bleach.3}
\end{eqnarray}
where the matrix $\Theta_{ab}$ is defined as
\begin{eqnarray}
\Theta_{ab} = \frac{1}{2} \phi_0 \delta_{ab} + \frac{1}{2} \epsilon_{abc} \phi_c \, .
\label{q.bleach.4}
\end{eqnarray}
At one loop order the LFE reads
\begin{eqnarray}
\Theta_{ab} \frac{\delta \G^{(1)}}{\delta \phi_b} = 0 \, .
\label{q.bleach.5}
\end{eqnarray}
Since the matrix $\Theta_{ab}$ is invertible the above
equation implies that 
 the dependence on the Goldstone fields is only via
the bleached variables. At higher orders one has to take
into account the inhomogeneous term in eq.(\ref{bracket.2}).
In addition to the dependence through the bleached
variables (implicit dependence), 
an additional explicit dependence of $\G^{(n)}$ on $\phi_a$ 
arises \cite{Bettinelli:2007kc}.
The integration can be explicitly carried out in an elegant way
by introducing the homotopy operator associated with
${\cal W}_0$, as discussed in \cite{Bettinelli:2007kc}.

\medskip
The bleached variables 
$w_{\mu}, \tilde L, \tilde {\bar L} $ as 
well as $R, \bar R$ and the $U(1)$ connection $B_\mu$ are both ${\cal W}_0$- and ${\cal S}_0$-invariant.
Moreover, by eq.(\ref{fle.5}) the ${\cal S}_0$-transforms of 
bleached variables are bleached. 

{The solution of the linearized STI can thus 
be studied in the space spanned by the bleached variables. 
Since the theory is non-anomalous, the
dependence on the bleached ghost $\tilde c$,
on the bleached anti-fields,
on the bleached background gauge source $v_\mu$ and 
its BRST partner ${\tilde \Theta}_\mu$ 
in eq.(\ref{q.bleach.1})  and on $\tilde K_0$ in 
eq.(\ref{q.bleach.2})
is confined
to the cohomologically trivial sector of ${\cal S}_0$-invariants
which are of the form ${\cal S}_0(X)$, where
$X$ is a local functional with ghost number $-1$ \cite{Henneaux:1998hq}.
}

{
This allows us to classify the possible invariant solutions 
by the same technique developed in \cite{Bettinelli:2007tq} for the $SU(2)$ case.
This strategy has been applied in order to obtain the complete set of one loop
counterterms  for the massive nonlinearly  realized $SU(2)$
Yang-Mills theory in \cite{Bettinelli:2007cy} .

We briefly illustrate the procedure at the one loop level  (the full algebraic analysis
is beyond the scope of the present paper and will be developed elsewhere).  
By the WPC formula in eq.(\ref{wpc}) the one-loop invariants can have at most dimension $4$.
According to the classification described above, they fall into two categories:
the first (cohomologically non-trivial sector) is spanned by the Lorentz-invariant 
electrically neutral monomials
in $w_\mu$, $\tilde L$, $\tilde {\bar L}$, $R$, $\bar R$ and ordinary derivatives thereof with
dimension $\leq 4$.  

The second class contains the cohomologically trivial electrically neutral invariants
with dimension $\leq 4$.
As an example, we write  the allowed cohomologically trivial invariants involving
the bleached anti-field $\widetilde {\hat A^*}_{a\mu}$ 
\begin{eqnarray}
&& {\cal J}_1 = \int d^Dx \, {\cal S}_0 (\widetilde {\hat A^*}_{a\mu} w_a^\mu)  ~ \, ,  ~~~~~ 
{\cal J}_2 = \int d^Dx \, {\cal S}_0 (\widetilde {\hat A^*}_{3\mu} w_3^\mu) \, , \nonumber \\
&& {\cal J}_3 = \int d^Dx \, {\cal S}_0 (\widetilde {\hat A^*}_{a\mu} v_a^\mu) ~ \,  , ~~~~~ 
{\cal J}_4 = \int d^Dx \, {\cal S}_0 (\widetilde {\hat A^*}_{3\mu} v_3^\mu ) \, .
\label{ex.1}
\end{eqnarray}
Notice that each invariant of the form $\int d^Dx \, {\cal S}_0 ( M_{ab} \widetilde {\hat A^*}_{a\mu} w_b^\mu)$
and $\int d^Dx \, {\cal S}_0 ( N_{ab} \widetilde {\hat A^*}_{a\mu} v_b^\mu)$,
with $M_{ab}, N_{ab}$ real matrices, would be allowed on the basis of the STI in eq.(\ref{brst.13}).
The requirement of invariance under the Abelian STI in eq.(\ref{st.3}) leaves only the four
invariants in eq.(\ref{ex.1}).
}

\subsection{Minimal Subtraction Procedure}\label{sec.subtr}

The superficial degree of divergence in eq. (\ref{wpc})
shows that the number of divergent amplitudes increases
order by order in the loop expansion, though it
remains finite at each order.
Therefore the theory is not power-counting renormalizable
even if we restrict to ancestor amplitudes.
This item has been considered at length by the
present authors. The extensive discussion is in
Ref. \cite{Bettinelli:2007zn}, where we argue
in favor of a particular {\em Ansatz} for the subtraction procedure
which respects locality and unitarity (at variance
with the algebraic renormalization which in the present
case leads to finite symmetric renormalizations which
cannot be reinserted back into the tree-level
vertex functional).
\par
In this approach eq. (\ref{bkgwi}) is used as a guide
in order to work out the procedure of the removal of 
divergences. Dimensional
regularization provides the most natural environment.
Let us denote by
\begin{eqnarray}
\G^{(n,k)}
\label{fle.6}
\end{eqnarray}
the vertex functional for 1-PI
amplitudes at $n$- order in loops where the
countertems enter with a total power $k$ in $\hbar$.
In dimensional regularization we can perform
a grading in $k$ of eq. (\ref{bkgwi}). Thus if
we have successfully performed the subtraction
procedure satisfying eq. (\ref{bkgwi}) up to
order $n-1$ the next order effective action
\begin{eqnarray}
\G^{(n)}=\sum_{k=0}^{n-1}\G^{(n,k)}
\label{fle.7}
\end{eqnarray}
violates eq. (\ref{bkgwi}) since the counterterm
$\hat \G^{(n)}$ is missing. The breaking term can be
determined by writing eq. (\ref{bkgwi}) at order 
$n$ at the grade $k\le n-1$ and then by summing over $k$.
One gets
\begin{eqnarray}
&&
{\cal W}_0\G^{(n)}
+ \frac{1}{2\Lambda^{(D-4)}}
\sum_{n'=1}^{n-1}\Bigl(
\frac{\delta\G^{(n-n')}}{\delta{K_0}}\Bigr)\Bigl(
\frac{\delta\G^{(n')}}{\delta{\phi_a}}\Bigr)
\nonumber\\&&
=\frac{1}{2\Lambda^{(D-4)}}
\sum_{n'=1}^{n-1}\Bigl(
\frac{\delta\G^{(n-n',n-n')}}{\delta{K_0}}\Bigr)\Bigl(
\frac{\delta\G^{(n',n')}}{\delta{\phi_a}}\Bigr).
\label{fle.8}
\end{eqnarray}
The first term in the l.h.s. of eq. (\ref{fle.8}) has pole parts in $D-4$
while the second is finite, since the factors
 are of order less than $n$, thus
already subtracted. The breaking term contains only counterterms
$\hat\G^{j}=\G^{(j,j)},~j<n$. This suggests the {Ansatz} that
the finite part of the Laurent expansion at $D=4$
\begin{eqnarray}
 \frac{1}{\Lambda^{(D-4)}}\G^{(n)}
\label{fle.9}
\end{eqnarray}
gives the correct prescription for the subtraction of the divergences;
i.e. one has to divide both members of eq. (\ref{fle.8}) by
$\Lambda^{(D-4)}$ and
remove only the pole parts (Minimal Subtraction). Thus the
counterterms have the form
\begin{eqnarray}
\hat\G^{(n)}= \Lambda^{(D-4)}\int \frac{d^Dx}{(2\pi)^D}{\cal M}^{(n)}(x)
\label{fle.10}
\end{eqnarray}
where the integrand is a local power series
in the fields, the external sources and their
derivatives (a local polynomial as far as
ancestor monomials are concerned) and it possesses only pole parts
in its Laurent expansion at $D=4$.

A similar argument applies to the STI in eq.(\ref{st.fid.1})
since the bracket in eq.(\ref{bracket.1}) has the same prefactor  
$\Lambda^{-(D-4)}$.
The $U(1)$ identity in eq.(\ref{charge.1}), being linear in $\G$,
does not pose any problem.
Compatibility of the STI and the LFE follows
from eq.(\ref{fle.5}). 
 
\par
In this subtraction scheme  one {extra} free parameter
enters, i.e. the overall mass scale $\Lambda$ for the
radiative corrections.

In this scheme the $\gamma_5$ problem is
treated in a pragmatic approach (for a similar
treatment see e.g. 
\cite{Jegerlehner:2000dz}). 
The matrix $\gamma_5$ is replaced by a 
new $\gamma_D$ which anti-commutes with
every $\gamma_\mu$. No statement is made
on the analytical properties of the traces involving
$\gamma_D$. Since the theory is not anomalous
such traces never meet poles in $D-4$ and therefore we
can evaluate at the end the traces at $D=4$. 

\par
In practice there are two ways to proceed in the regularization
procedure. One can use the forest formula and use Minimal Subtraction
for every (properly normalized) subgraph. It is possible, as alternative,
to evaluate the counterterms for the ancestor amplitudes and
then obtain from those all the necessary counterterms involving
the Goldstone boson fields~$\vec\phi$. 

\section{The Neutral Sector}
\label{sec:neu}
{
The existence of two equations (STI and LFE), together with the LGE, allows
to derive a surprisingly rich set of results for the neutral
sector. We focus on those that are relevant for the identification
of the photon field after radiative corrections. In this Section
and in the attached Appendix \ref{app:neutral} we use  a simplified
notation
\begin{eqnarray}&&
W_{A_1\cdots A_n}= \frac{\delta^n W}{\delta J_{A_1}\cdots \delta J_{A_n}}=i^{(n-1)} 
\langle 0|T(A_1\cdots A_n)|0\rangle
\label{neu.1}
\end{eqnarray}
where $J_{A_1}$ is the source for $A_1$. Moreover we use the conventions
\begin{eqnarray}
{\mathfrak M}_A
\label{neu.2}
\end{eqnarray}
for an S-matrix element on which the functional derivative
with respect to $J_A$ has been taken and all external sources 
have been put to zero. The states resulting from the reduction formulas are
not displayed, if not necessary. Finally the $\widehat A$ indicates that
the external leg attached to $A$ has been removed. For instance
\begin{eqnarray}
{\mathfrak M}_{\widehat A}.
\label{neu.2.1}
\end{eqnarray}
\par
By taking the appropriate linear combination
of the Abelian STI in eq.(\ref{st.3})  and
the third component of the LFE in eq.(\ref{bkgwi}), 
the bilinear term $\G_{K_0} \G_{\phi_3}$ can be removed. This yields
\begin{eqnarray}&& 
-{\frac{1}{g'}} \Lambda^{(D-4)}\Box b_0
+\Biggl(
- {\frac{1}{g'}}\partial^\mu  \frac{\delta }{\delta  B^\mu}
-{\frac{1}{g}}\partial_\mu \frac{\delta }{\delta A_{3 \mu}} 
-{\frac{1}{g}}\partial_\mu \frac{\delta }{\delta V_{3 \mu}}
\nonumber \\&&  
+ A_{2\mu} \frac{\delta }{\delta A_{1\mu}} 
-A_{1\mu} \frac{\delta }{\delta A_{2\mu}}
+iQL\frac{\delta }{\delta L}
-i\bar L Q\frac{\delta }{\delta \bar L}
+i Q R\frac{\delta }{\delta R}
-i\bar R Q\frac{\delta }{\delta \bar R}
\nonumber \\&&  
+ \phi_2 \frac{\delta }{\delta \phi_1}
-\phi_1\frac{\delta }{\delta \phi_2}
+  b_2 \frac{\delta }{\delta b_1}- b_1 \frac{\delta }{\delta b_2}
      +  c_2 \frac{\delta }{\delta  c_1}
      -  c_1 \frac{\delta }{\delta  c_2}
\nonumber \\&&  
      +  \bar c_2 \frac{\delta }{\delta \bar c_1}
      -  \bar c_1 \frac{\delta }{\delta \bar c_2} 
+  V_{2\mu} \frac{\delta }{\delta V_{1\mu}}
-  V_{1\mu} \frac{\delta }{\delta V_{2\mu}}
+ \Theta_{2\mu} \frac{\delta }{\delta \Theta_{1\mu} }
- \Theta_{1\mu} \frac{\delta }{\delta \Theta_{2\mu} }
\nonumber \\&&  
      + A^*_{2\mu} \frac{\delta }{\delta A^*_{1\mu}}
      - A^*_{1\mu} \frac{\delta }{\delta A^*_{2\mu}}
+ \phi_2^* \frac{\delta }{\delta \phi_1^*}
-\phi_1^* \frac{\delta }{\delta \phi_2^*}
      +  c^*_2 \frac{\delta }{\delta  c^*_1} 
      -  c^*_1 \frac{\delta }{\delta  c^*_2} 
\nonumber\\&&
-iQL^*\frac{\delta }{\delta L^*}
+i\bar L^*Q\frac{\delta }{\delta \bar L^*}
\Biggr)\G
= 0 \, ,
\label{charge.1}
\end{eqnarray}
where $Q$ is the electric charge of the component of the
multiplet. In terms of the fields
\begin{eqnarray}&& 
Z_\mu = \frac{1}{\sqrt{g^2+g^{'2}}}(g A_{3\mu}-g' B_\mu)
\nonumber\\&&
A_\mu = \frac{1}{\sqrt{g^2+g^{'2}}}(g' A_{3\mu}+g B_\mu),
\label{charge.1.1}
\end{eqnarray}
the neutral boson part in eq. (\ref{charge.1}) takes the form
\begin{eqnarray}
- \frac{1}{g'}\partial^\mu  \frac{\delta }{\delta  B^\mu}
-\frac{1}{g}\partial_\mu \frac{\delta }{\delta A_{3 \mu}}
= - \frac{\sqrt{g^2+g^{'2}}}{gg'} \partial^\mu  \frac{\delta }{\delta  A^\mu}.
\label{charge.1.2}
\end{eqnarray}
The term $-\frac{1}{g}\partial_\mu \frac{\delta }{\delta V_{3 \mu}}$
takes into account that the fields of the photon and of the $Z_0$, as superposition
of $(A_{3\mu},B_\mu)$, are modified by the
perturbative corrections. 
\par
In the generic S matrix elements  the insertion of $V_{a\mu}$ 
is  zero for physical states. The proof makes use of the STI in eq.
(\ref{brst.13})
written for the connected amplitude
\begin{eqnarray}
&& \!\!\!\!\!\!\!\!\!\!
{\cal S}W \equiv \int d^Dx \,\Biggl[ \Lambda^{-(D-4)}\Big (
- W_{ A^*_{a\mu}}  J_{a\mu}
-
 W_{ \phi_a^*} K_a
+ 
 W_{ c_a^*}\bar \eta_a
 \nonumber \\
&&
+ W_{ L^*} \bar\xi
+ W_{ \bar L^*} \xi \Bigr)
+ \eta_a W_{b_a}
 + \Theta_{a\mu}  W_{ V_{a\mu}}
      - K_0  W_{ \phi_0^*} 
\Biggr] = 0 
\label{brst.13.1}
\end{eqnarray}
where $ J_{a\mu}A_a^\mu+K_a\phi_a+\bar \eta_a c_a+ \bar c_a\eta_a+
\bar L\xi+\bar\xi L$ are the source terms. One takes
the functional derivative with respect to $\Theta_{a\mu}$ and
subsequently applies the procedure of deriving with respect to the
field sources and finally applies the reduction formulas. On the physical
states one obtains
\begin{eqnarray} 
 {\mathfrak M}_{ V_{a\mu}\dots}=0
\label{brst.13.2}
\end{eqnarray}
where the dots $\dots$ indicate the physical state variables. 
Consequently from 
the WTI (\ref{charge.1})
(written for the connected amplitudes) we get 
\begin{eqnarray} 
\Box  {\mathfrak M}_{ b_0\dots}=0.
\label{neu.3}
\end{eqnarray}
A further important identity can be derived from eq. (\ref{brst.13.1}). By 
differentiating with respect to $\eta_3$ and by constructing a physical
S-matrix element, one gets
\begin{eqnarray} 
 {\mathfrak M}_{ b_3\dots}=0.
\label{neu.3.1}
\end{eqnarray}
%

\subsection{The two-point Functions}
In this subsection we determine the most general form of the two-point
functions in the Landau gauge. For this purpose we use the STI, LFE, and LGE, where we drop  
 all the terms that cannot produce any contributions. 
Moreover we impose the condition
\begin{eqnarray}
\Gamma\cdot W= - I\!\!I.
\label{neu.0}
\end{eqnarray}
The explicit calculation is given in Appendix \ref{app:neutral} and the results
can be displayed in a matrix form both for $\Gamma$ and $W$.

}


Summary for the two-point function  $\Gamma$
\begin{equation}
\left(
\begin{array}{lllrrr}
&A_3^\mu &  B^\mu & b_3&b_0 & \phi_3 \\
\\
A_3^\nu & \Gamma_T^{AA}
{\scriptstyle{\prod}}^{\mu\nu}+ \Gamma_L^{AA}\frac{p^\mu p^\nu}{p^2} 
&\Gamma_T^{AB}{\scriptstyle{\prod}}^{\mu\nu}
+ \Gamma_L^{AB}\frac{p^\mu p^\nu}{p^2}  
& i \Lambda^{D-4}p^\nu &0&- i \frac{ 2 p^\nu}{ v' g'}\Gamma_L^{AB} \\
B^\nu & \Gamma_T^{BA}
{\scriptstyle{\prod}}^{\mu\nu}+ \Gamma_L^{BA}\frac{p^\mu p^\nu}{p^2}
&\Gamma_T^{BB}
{\scriptstyle{\prod}}^{\mu\nu}+ \Gamma_L^{BB}\frac{p^\mu p^\nu}{p^2} &0
&i \Lambda^{D-4}p^\nu
&-i \frac{2 p^\nu}{v' g'}\Gamma_L^{BB}\\
b_3& -i\Lambda^{D-4}p^\mu &0& 0 & 0& 0\\
b_0&0&-i \Lambda^{D-4}p^\mu &0&0&0\\
\phi_3& i \frac{ 2 p^\mu}{ v' g'}\Gamma_L^{BA}  
 &i \frac{2 p^\mu}{v' g'}\Gamma_L^{BB}& 0& 0
&p^2 \biggl(\frac{2}{v' g'}\biggr)^2\Gamma_L^{BB}
\end{array}
\right),
\label{fey.7}
\end{equation}
Summary for the propagator $W$
\begin{equation}
\left(
\begin{array}{lllrrr}
&A_3^\mu &  B^\mu & b_3&b_0 & \phi_3 \\
\\
A_3^\nu & -\frac{\Gamma_T^{BB}}{\Delta_T}
{\scriptstyle{\prod}}^{\mu\nu}&\frac{\Gamma_T^{AB}}{\Delta_T}
{\scriptstyle{\prod}}^{\mu\nu}& -i\frac{p^\nu}{\Lambda^{D-4}p^2}&0&0 \\
B^\nu &\frac{\Gamma_T^{AB}}{\Delta_T} 
{\scriptstyle{\prod}}^{\mu\nu}&-\frac{\Gamma_T^{AA}}{\Delta_T}
{\scriptstyle{\prod}}^{\mu\nu}& 0&-i\frac{p^\nu}{\Lambda^{D-4}p^2}&0 \\
b_3& i\frac{p^\mu}{\Lambda^{D-4}p^2}&0& 0& 0&
-\frac{v' g' \Gamma_L^{AB}}{2\Lambda^{D-4}p^2 \Gamma_L^{BB}}\\
b_0&0&i\frac{p^\mu}{\Lambda^{D-4}p^2}&0&0&-  \frac{v' g'}{2p^2 \Lambda^{D-4}}\\
\phi_3&0&0&-\frac{v' g' \Gamma_L^{AB}}{2\Lambda^{D-4}p^2 \Gamma_L^{BB}}&
 -  \frac{v' g'}{2p^2\Lambda^{D-4}}
&W_{\phi_3\phi_3}
\end{array}
\right),
\label{fey.8}
\end{equation}
where
\begin{eqnarray}
\Delta_T= \Gamma_T^{AA}\Gamma_T^{BB}-\Gamma_T^{AB}\Gamma_T^{BA},\quad
{\scriptstyle{\prod}}^{\mu\nu}= g^{\mu\nu}- \frac{p^\mu p^\nu}{p^2},
\quad v' = \Lambda^{-D+4}\Gamma_{K_0}.
\label{charge.46}
\end{eqnarray}
We see from eq. (\ref{fey.7}) that the field
\begin{eqnarray}
A^\mu \equiv \frac{1}{\sqrt{ (\Gamma_L^{BB})^2+(\Gamma_L^{AB})^2}}
\biggl(
-\Gamma_L^{BB} A_3^\mu + \Gamma_L^{AB} B^\mu
\biggr)
\label{charge.47.1}
\end{eqnarray}
decouples from $\phi_3$. While the corresponding orthogonal combination 
\begin{eqnarray}
Z^\mu \equiv \frac{1}{\sqrt{ (\Gamma_L^{BB})^2+(\Gamma_L^{AB})^2}}
\biggl(
\Gamma_L^{AB} A_3^\mu + \Gamma_L^{BB} B^\mu
\biggr)
\label{charge.47.3}
\end{eqnarray}
remains coupled to  $\phi_3$. Moreover again from  eq. (\ref{fey.7})
we see that the longitudinal part of the 1-PI two-point function of $A_\mu$ 
is zero while it
remains non zero for $Z_\mu$. This is due to the fact that $\Delta_L=0$.
In fact from Appendix \ref{app:neutral} eq. (\ref{charge.48.2})  we have
\begin{eqnarray}
\frac{\Gamma_L^{AA}}{\Gamma_L^{AB}}= \frac{\Gamma_L^{BA}}{\Gamma_L^{BB}}
=-\frac{2p^2} {v'g'}\frac{\Gamma_{c_3\phi_3^*}} {\Gamma_{c_3\bar c_3}}.
\label{neu.50}
\end{eqnarray}
The above equation (\ref{neu.50}) shows also that the Nakanishi-Lautrup 
Lagrange multiplier for $A_\mu$
\begin{eqnarray}
b_A \equiv \frac{1}{\sqrt{ (\Gamma_L^{BB})^2+(\Gamma_L^{AB})^2}}
\biggl(
-\Gamma_L^{BB} b_3 + \Gamma_L^{AB} b_0
\biggr)
\label{charge.47.2p}
\end{eqnarray}
decouples from $\phi_3$. 

\subsection{Decoupling of the Unphysical Modes in the Neutral Sector at $p^2=0$}

At $p^2=0$ there are some unphysical modes in the neutral sector. 
They show up in the propagator of the $Z_\mu$ in the Landau gauge and 
in the propagator of the $\phi_3$. There is a further $p^2=0$ unphysical
pole in the photon propagator. 
\par
We have eq. (\ref{neu.3}) which follows from the WTI (\ref{charge.1}) and
(\ref{brst.13.2}). In the limit $p^2=0$ only the pole parts survive.
By using the relations in eq. (\ref{fey.8}) 
the WTI (\ref{neu.3}) yields
\begin{eqnarray}
\Bigl[ip^\mu {\mathfrak M}_{\widehat{B^\mu}\cdots} 
- \frac{v' g'}{2}{\mathfrak M}_{\widehat{\phi_3}\cdots}\Bigr]_{p^2=0}=0.
\label{charge.5}
\end{eqnarray}
Now we use  eq. (\ref{neu.3.1}). The  multiplication
by the square of the external momentum and its limits to zero selects
only the pole parts. From eqs. (\ref{neu.5}), (\ref{neu.7}) and
(\ref{neu.10}) in Appendix \ref{app:neutral} and eq. (\ref{neu.50})
\begin{eqnarray}
\lim_{p^2=0}p^2{\mathfrak M}_{b_3\cdots}=
\lim_{p^2=0}\biggl(i p^\mu {\mathfrak M}_{\widehat{A_3^\mu}\cdots}
-\frac{v' g'}{2} \frac{\Gamma_L^{AB}}{\Gamma_L^{BB}}{\mathfrak M}_{\widehat{\phi_3}\cdots}
\biggr)=0.
\label{charge.49}
\end{eqnarray}
By removing the contribution of $\phi_3$ between eqs. (\ref{charge.5}) and 
(\ref{charge.49}) we get
\begin{eqnarray}
\lim_{p^2=0}p^\mu\biggl(  {\mathfrak M}_{\widehat{A_3^\mu}\cdots}
- \frac{\Gamma_L^{AB}}{\Gamma_L^{BB}}{\mathfrak M}_{\widehat{B^\mu}\cdots}
\biggr)=0
\label{charge.50}
\end{eqnarray}
which guarantees that longitudinally polarized photons decouple
from physical states.
Now we consider the massless modes present in the $Z_\mu$ sector.
The combination of eqs. (\ref{charge.5})  and 
(\ref{charge.49}) orthogonal to the
one in eq. (\ref{charge.50}) is
\begin{eqnarray}
\lim_{p^2=0}\biggl(i p^\mu \Gamma_L^{AB}
{\mathfrak M}_{\widehat{A_3^\mu}\cdots}+i p^\mu \Gamma_L^{BB}
{\mathfrak M}_{\widehat{B^\mu}\cdots}
- \frac{v' g'}{2} \frac{(\Gamma_L^{AB})^2+(\Gamma_L^{BB})^2 }{\Gamma_L^{BB}}{\mathfrak M}_{\widehat{\phi_3}\cdots}
\biggr)=0.
\label{charge.51}
\end{eqnarray}
The $Z-Z$ propagator (\ref{fey.8}) written for the linear combination
 (\ref{charge.47.3}) is
\begin{eqnarray}
W_{Z^\mu Z^\nu} = 
\frac{{\scriptstyle{\prod}}^{\mu\nu}}{\Delta_T[(\Gamma_L^{AB})^2+(\Gamma_L^{BB})^2]}
\biggl(- \Gamma_T^{BB}\Gamma_L^{AB}\Gamma_L^{AB}+2\Gamma_T^{AB}\Gamma_L^{BB}\Gamma_L^{AB}
- \Gamma_T^{AA}\Gamma_L^{BB}\Gamma_L^{BB}
\biggr).
\label{charge.52}
\end{eqnarray}
Now we require that the two-point functions $\Gamma $ be non singular at $p^2=0$, i.e. \cite{Ferrari:2004pd}
\begin{eqnarray}
\lim_{p^2=0}(\Gamma_T^{XY}-\Gamma_L^{XY})=0,
\label{charge.53}
\end{eqnarray}
\begin{eqnarray}
W_{Z^\mu Z^\nu}|_{p^2\sim 0} = 
\frac{{\scriptstyle{\prod}}^{\mu\nu}}{[(\Gamma_L^{AB})^2+(\Gamma_L^{BB})^2]}
\biggl(- \Gamma_L^{BB}
\biggr).
\label{charge.54}
\end{eqnarray}
%
Eqs. (\ref{charge.51}) and (\ref{charge.54}) imply 
\begin{eqnarray}&&
\lim_{p^2=0} p^2{\mathfrak M}_{\widehat{Z_\mu}\cdots}^*W_{Z^\mu Z^\nu}{\mathfrak M}_{\widehat{Z_\nu}\cdots}
= \lim_{p^2=0} 
{\mathfrak M}_{\widehat{\phi_3}\cdots}^* \frac{(g' v')^2}{4\Gamma_L^{BB}}{\mathfrak M}_{\widehat{\phi_3}\cdots}
\nonumber\\&&
= -\lim_{p^2=0} p^2
{\mathfrak M}_{\widehat{\phi_3}\cdots}^* W_{\phi_3\phi_3}
{\mathfrak M}_{\widehat{\phi_3}\cdots}.
\label{charge.55}
\end{eqnarray}
The last term cancels the $\phi_3$ contribution coming from the full propagator (\ref{fey.8}).


\section{Conclusions}\label{sec.concl}

The electroweak model based on the nonlinearly realized
$SU(2) \otimes U(1)$ gauge group can be consistently 
defined in the perturbative loop-wise expansion.
In this formulation there is no Higgs in the perturbative series.

The present approach is based on the LFE and the WPC.
The LFE encodes the invariance of the path-integral
Haar measure under local $SU(2)_L$ transformations
and provides a hierarchy among 1-PI Green functions
by fixing all amplitudes involving at least one Goldstone leg.
The ancestor amplitudes (i.e. those with no Goldstone legs)
obey the WPC theorem.

There is a unique classical action giving rise to Feynman rules
compatible with the WPC formula in eq.(\ref{wpc}).
In particular the anomalous couplings, which would be otherwise allowed
on symmetry grounds, are excluded by the WPC. 
Two gauge boson mass invariants
are compatible with the WPC and the symmetries. Thus the
tree-level Weinberg relation is not working in the nonlinear
framework.

The discovery of the LFE suggests a unique 
{\em Ansatz} for the subtraction
procedure which is symmetric, i.e it respects the LFE
itself, the STI (necessary for the fulfillment of the
Physical Unitarity) and the LGE (controlling the
stability of the gauge-fixing under radiative corrections).
A linear Ward identity exists for the electric charge
(despite the nonlinear realization of the gauge group).
The strategy does not alter the 
number of tree-level parameters apart from a common mass
scale of the radiative corrections. The algorithm 
is strictly connected with dimensional regularization and the symmetric
subtraction of the pole parts in the Laurent expansion of the 1-PI
amplitudes.

The theoretical and phenomenological consequences of this scenario
are rather intriguing. 
A Higgs boson could emerge as a non-perturbative 
mechanism, but then its physical parameters are not
constrained by the radiative corrections of the low energy
electroweak processes. Otherwise the energy scale
for the radiative corrections $\Lambda$ is a manifestation
of some other high-energy physics.

Many aspects remain to be further studied. We only mention some
of them here.
The issue of unitarity at large energy (violation of Froissart bound) 
\cite{Froissart:1961ux} at fixed order
in perturbation theory when the Higgs field is removed 
(as in \cite{Lee:1977eg},\cite{Longhitano:1980iz},\cite{Longhitano:1980tm})  
can provide additional insight in the role of the mass scale $\Lambda$. 
The electroweak model
based on the nonlinearly realized gauge group
 satisfies Physical Unitarity as a consequence
of the validity of the Slavnov-Taylor identity. Therefore 
violation of the Froissart bound can only occur in evaluating cross sections
at finite order in perturbation theory. This requires the evaluation 
of a scale at each order where unitarity at large energy is substantially violated.

The phenomenological implications of the nonlinear theory
in the electroweak precision fit have to be investigated.

Finally the extension of the present approach to
larger gauge groups
(as in Grand-Unified models)  could help in
understanding  the nonlinearly realized spontaneous
symmetry breaking  mechanism (selection
of the identity as the preferred direction in the $SU(2)$ manifold)  and the 
associated appearance of two independent gauge boson mass invariants.

\section*{Acknowledgements} 
{
One of us (R.F) is pleased to thank the Center for Theoretical Physics at MIT,
Massachusetts, where he had the possibility to work on this research.}
{A.Q.} would like to thank the Max-Planck-Institut f\"ur Physik in Munich
and the Institut f\"ur Kernphysik at the Technische Hochschule Dresden for the
warm hospitality. Useful discussions with S.~Dittmaier and D.~St\"ockinger are
gratefully acknowledged.

\appendix

\section{Propagators in the Landau gauge}\label{app.prop}

We summarize here the propagators in the Landau gauge
evaluated in the symmetric formalism. 
It is convenient to rescale the Goldstone fields according to
\begin{eqnarray}
\phi_{1,2} \rightarrow \frac{v}{2M} \phi_{1,2} \, , ~~~~
\phi_3 \rightarrow \frac{v}{2M(1+\kappa)^{1/2}} \, .
\label{prop.0} 
\end{eqnarray}
This ensures the common normalization of the 
Goldstone propagators.
We  define the Weinberg angle via the relation
\begin{eqnarray}
\tan \theta_W = \frac{g'}{g} \, .
\label{prop.0.1}
\end{eqnarray}
The sine and cosine of the Weinberg angle are denoted by
\begin{eqnarray}
c = \cos \theta_W \, , ~~~~ s = \sin \theta_W \, .
\label{prop.0.2}
\end{eqnarray}
We also define the masses of the charged and neutral
gauge boson mass eigenstates:
\begin{eqnarray}
M^2_W = (gM)^2 \,  , ~~~~ M^2_Z = \frac{(gM)^2}{c} (1 + \kappa) \, ,
\label{prop.0.3}
\end{eqnarray}
By inverting the two-point functions in $\G^{(0)}$ 
in eq.(\ref{new.1}) one finds
(the common pre-factor 
$\Lambda^{-(D-4)}$ is always left understood)
\begin{eqnarray}
&& \Delta_{A_{1\mu} A_{1\nu}} = \Delta_{A_{2\mu} A_{2\nu}} = 
\frac{i}{-p^2 + M_W^2} T_{\mu\nu} \, , ~~~~
\Delta_{A_{1\mu}A_{2\nu}}=  \Delta_{A_{1\mu} A_{3\nu}} = \Delta_{A_{2\mu} A_{3\nu}} = 0 \, , 
\nonumber \\
&& \Delta_{A_{3\mu} A_{3\nu}} = 
\frac{i}{-p^2 + M_Z^2} T_{\mu\nu} \, , ~~~~
\Delta_{A_{3\mu} B_\nu} = cs \Big ( \frac{i}{-p^2} -
\frac{i}{-p^2+ M_Z^2} \Big ) T_{\mu\nu} \, , \nonumber \\
&& \Delta_{A_{1\mu} B_\nu} = \Delta_{A_{2\mu} B_\nu} = 0 \, , 
~~~~ \Delta_{B_\mu B_\nu} = \Big ( c^2 \frac{i}{-p^2}
+ s^2 \frac{i}{-p^2 + M_Z^2} \Big ) T_{\mu\nu}\, , 
\nonumber \\
&& 
\Delta_{\phi_a \phi_b} = \delta_{ab} \frac{i}{p^2} \, ,
\nonumber \\
&& 
\Delta_{b_i A_{j\mu}} = -  \frac{p_\mu}{p^2} \delta_{ij} \, , ~~~~
\Delta_{b_i b_j} = 0 \, , ~~~~ \Delta_{b_i \phi_j} = -i \delta_{ij} 
 \frac{M_W}{p^2} \, , ~~~~ i,j =1,2 \, , 
\nonumber \\
&&
\Delta_{B_3 A_{3\mu}} = \frac{- p^\mu}{p^2} \, , ~~~~
\Delta_{B_3 B_\mu} = 0 \, ,  ~~~~~~
\Delta_{B_3 \phi_3} = - i c \frac{M_Z}{p^2} \, ,  ~~ 
\Delta_{B_3 B_3} = 0 \, , \nonumber \\
&& \Delta_{b_0 A_{3\mu}} =  0 \, , ~~~~~~~~ 
\Delta_{b_0 B_\mu} =  -\frac{ p^\mu}{p^2} \, , ~~~~ 
\Delta_{b_0 \phi_3} = i s \frac{M_Z}{p^2} \, , ~~~~~ 
\Delta_{b_0 b_0} = 0 \, , \nonumber \\
&&
\Delta_{\bar c_a c_b}  = \delta_{ab} \frac{i}{p^2} \, , ~~~~
\Delta_{\bar c_0 c_0} = \frac{i}{p^2} \, . 
\label{prop.2}
\end{eqnarray}
The mixed $A-\phi$ propagators are zero.

The relation with the mass eigenstates is given by
\begin{eqnarray}
&& A_\mu = c B_\mu + s A_{3\mu} \, , ~~~~~~ Z_\mu = -s B_\mu + c A_{3\mu} \, .
\label{prop.3}
\end{eqnarray}

In the fermion sector the propagators are 
\begin{eqnarray}
\Delta_{\bar f f} = \frac{i}{\slshp - m_f} 
\label{prop.4}
\end{eqnarray}
where $m_f$ is the mass of the fermionic species $f$.

\section{Proof of the Weak-Power Counting Formula}
\label{app:C}

In this Appendix we prove the weak power-counting formula in 
eq.(\ref{wpc}) by extending 
the analysis carried out for 
massive $SU(2)$ Yang-Mills theory  \cite{Bettinelli:2007tq}
to the electroweak model
based on the nonlinearly realized $SU(2)_L \otimes U(1)_R$
gauge group.

Let ${\cal G}$ 
be an arbitrary $n$-loop 1-PI ancestor graph with $I$ internal lines, $V$ vertexes and a given set $\{ N_A, N_B, N_F, N_{\bar F}, N_c, N_V, N_\Omega, N_{\phi_0^*},
N_{K_0}, N_{\phi_a^*}, N_{A^*}, N_{c^*}, N_{L^*}, N_{\bar L^*} \}$ of external legs.
$F,\bar F$ are a collective notation for the fermion and anti-fermion
matter fields, which can be treated in a unified manner.

We do not need to consider $\Delta_{b_0B}$ since there
are no vertexes involving $b_0$.
By eq.(\ref{prop.2}) all the remaining propagators 
behave  as $p^{-2}$ as $p$ goes to infinity,
with the exception of $\Delta_{bA} \sim p^{-1}$.

Let us denote by
${\hat I}$ the number of internal lines associated
with propagators behaving as $p^{-2}$,
by $I_b$ the number of 
internal lines with propagators 
$\Delta_{bA}$
and by $I_F$ the number of internal fermionic lines. 
One has 
\begin{eqnarray}
I= {\hat I} + I_b +  I_F \, .
\label{wk.new}
\end{eqnarray}

According to the Feynman rules generated by
the tree-level vertex functional in eq.(\ref{new.1})
the superficial degree of divergence of ${\cal G}$ is 
\begin{eqnarray}
&& \!\!\!\!\!\!\!\!\!\!\!\!\!\!\!\!\!\!\!\!\!\!
d({\cal G}) =  n D - 2 {\hat I} - I_b -I_F + V_{AAA} \nonumber \\
&&           + \sum_k V_{A \phi^k} +
                     \sum_k V_{B\phi^k} +
                     2 \sum_k V_{\phi^k} + V_{\bar c c A}
                                         + V_{\bar c c V} \, .
\label{wk.2}
\end{eqnarray}
In the above equation we have denoted by $V_{AAA}$ the number of 
vertexes in ${\cal G}$ with three $A$-fields, 
with $V_{A\phi^k}$ the number of vertexes with one $A$
and $k$ $\phi$'s and so on.
By using eq.(\ref{wk.new}) we can rewrite eq.(\ref{wk.2}) as
\begin{eqnarray}
\!\!\!\!\!\!\!\!\!\!\!\!\!\!\!\!
d({\cal G}) & = & n D - 2 I + I_b +I_F + V_{AAA} \nonumber \\
&& + \sum_k V_{A \phi^k} +  \sum_k V_{B \phi^k} +
                     2 \sum_k V_{\phi^k} + V_{\bar c c A}
                                         + V_{\bar c c V} \, .
\label{wk.2.bis}
\end{eqnarray}
The total number of vertexes $V$ is given by
\begin{eqnarray}
V & = & V_{AAA} + V_{AAAA} + \sum_k V_{A \phi^k} 
+ \sum_k V_{B\phi^k} + \sum_k V_{\phi^k} \nonumber \\
  &   & +V_{bVA} + V_{\bar c c A} + V_{\bar c c V}
        +V_{\bar c c V A} \nonumber \\
  &   & + V_{\bar c A \Theta} +  V_{\phi_0^* \phi c}
        + \sum_k V_{\phi_a^* \phi^k c} \nonumber \\
  &   & + V_{A^* A c} + V_{c^* c c} 
        + \sum_k V_{K_0 \phi^k} \nonumber \\
   &   & + V_{\bar FFA} + \sum_k V_{\bar F F \phi^k} +
               V_{\bar F F B} + V_{{\bar L}^* \bar L c} +
               V_{L^* L c} \, .
\label{wk.3}
\end{eqnarray}

Euler's formula yields
\begin{eqnarray}
I = n + V -1 \, .
\label{wk.4}
\end{eqnarray}
Moreover, since $b$ only enters into the trilinear
vertex $\G^{(0)}_{b_a V_{b\mu} A_{c\nu}}$, 
the number
of $bVA$ vertexes must be greater than or equal to the number of propagators $\Delta_{bA}$
\begin{eqnarray} 
I_b \leq V_{bVA} \, .
\label{wk.2.ter}
\end{eqnarray}
On the other hand, the number of internal fermion lines
fulfills the following bound
\begin{eqnarray}
I_F \leq V_{\bar F FA} + V_{\bar F F B} + \sum_k V_{\bar F F \phi^k} \, .
\label{wk.ferm.1}
\end{eqnarray}

By using eqs.(\ref{wk.3}),(\ref{wk.4}), (\ref{wk.2.ter}) 
and (\ref{wk.ferm.1})
into eq.(\ref{wk.2}) one
gets
\begin{eqnarray}
d({\cal G}) & = & (D-2)n +2 + I_b + I_F \nonumber \\
     &   & - V_{AAA} - \sum_k V_{A\phi^k} 
              - \sum_k V_{B\phi^k}-V_{\bar c c  A}
           - V_{\bar c c V} \nonumber \\
     &   & - 2 \Big [ V_{AAAA} + V_{bVA} + V_{\bar c c V A}
                     +V_{\bar c A \Theta} \nonumber \\
     &   & ~~ + V_{\phi_0^* \phi c}
                   + \sum_k V_{\phi_a^* \phi^k c} 
                   + V_{A^* A c} + V_{c^* c c } 
                   + \sum_k V_{K_0 \phi^k} \nonumber \\
      &   & ~~ + V_{\bar F FA} + V_{\bar F F B} + \sum_k V_{\bar F F \phi^k}  + V_{{\bar L}^* \bar L c} +
               V_{L^* L c} \  \Big ] 
\nonumber \\
     & \leq & (D-2)n +2  \nonumber \\
     &   & - V_{AAA} - \sum_k V_{A\phi^k} 
         - \sum_k V_{B\phi^k}  -V_{\bar c c  A}
           - V_{\bar c c V} \nonumber \\
     &   & - V_{bVA} - V_{\bar F FA} - V_{\bar F F B} - \sum_k V_{\bar F F \phi^k}  \nonumber \\
     &   &- 2 \Big [ V_{AAAA} + V_{\bar c c V A}
                     +V_{\bar c A \Theta} \nonumber \\
     &   & ~~ + V_{\phi_0^* \phi c}
                   + \sum_k V_{\phi_a^* \phi^k c} 
                   + V_{A^* A c} + V_{c^* c c } 
                   + \sum_k V_{K_0 \phi^k} \nonumber \\
      &  & ~~ + V_{{\bar L}^* \bar L c} +
               V_{L^* L c} \Big ] \, .
\label{wk.5}
\end{eqnarray}

Clearly one has
\begin{eqnarray}
&& V_{\bar c A \Theta} = N_{\Theta} \, , ~~~~ 
V_{\phi_0^* \phi c} = N_{\phi_0^*} \, , 
\nonumber \\
&& V_{A^* A c} = N_{A^*} \, , ~~~~
   V_{c^* c c} = N_{c^*} \, , \nonumber \\
&& \sum_k V_{\phi_a^* \phi^k c} = N_{\phi_a^*} \, , ~~~~
\sum_k V_{K_0 \phi^k} = N_{K_0} \, , \nonumber \\
&& V_{\bar c c V} +  V_{bVA} +  V_{\bar c c V A} = 
 N_V \, ,  \nonumber \\
 && V_{{\bar L}^* \bar L c} = N_{{\bar L}^*} \, , ~~~~
       V_{L^* L c} = N_{L^*} \, .
\label{wk.6}
\end{eqnarray}
Moreover
\begin{eqnarray}
&& V_{AAA} + \sum_k V_{A \phi^k} + 2 V_{AAAA} + 
V_{\bar F F A}   
+ \sum_k V_{B \phi^k} + V_{\bar F F B} + \sum_k V_{\bar F F \phi^k}\nonumber \\
&& ~~~~~~~~
 + V_{\bar c c A}+  V_{\bar c c V A} + \sum_k V_{\phi_a^* \phi^k c}  \geq N_A + N_B + N_c  + N_F + N_{\bar F} \, .
\label{wk.8}
\end{eqnarray}
In fact the quadrilinear vertex $V_{AAAA}$ can give one 
or two external $A$ lines and the vertexes
$V_{\bar F F B}, V_{\bar F F A}$ can give rise to at most
one external $B$- and $A$- line respectively.

By using eqs.(\ref{wk.6}) and (\ref{wk.8}) into
eq.(\ref{wk.5}) we obtain 
in a straightforward way the following bound:
\begin{eqnarray}
&& d({\cal G}) \leq (D-2)n +2 - N_A - N_B - N_c - N_F 
- N_{\bar F}  -  N_V - N_{\phi_a^*} \nonumber \\
&& ~~~~ - 2 (N_\Theta + N_{A^*} + N_{\phi_0^*} + N_{L^*} + N_{{\bar L}^*} + N_{c^*} + N_{K_0} ) \, .
\label{wk.9.bis}
\end{eqnarray}
This establishes the validity of the weak power-counting formula.
\section{Two-point Functions Results}
\label{app:neutral}
The results of this Appendix are valid for a generic value of $p^2$. 
\par
From the $U(1)$ LGE
\begin{eqnarray}
-J_{b_0}= \Lambda^{(D-4)}\partial^\mu W_{B^\mu}
\label{neu.4}
\end{eqnarray}
we get
\begin{eqnarray}
&&
\!\!\!\!\!\!\!\!\!\!\!\!\!\!\!\!\!\!\!\!\!\!\!\!\!\!\!
W_{ B^\mu b_0} =- i\frac{ p_\mu}{\Lambda^{(D-4)}p^2},\quad W_{ B^\mu b_3} =0,\quad W_{ B^\mu \phi_3} =0
,\quad p^\mu W_{ B^\mu A_3^\nu} =0,\quad p^\mu W_{ B^\mu B^\nu} =0.
\label{neu.5}
\end{eqnarray}
From the $SU(2)$ LGE (\ref{b.eq})
\begin{eqnarray}
-J_{b_3}= \Lambda^{(D-4)}\partial^\mu( W_{A_3^\mu}-V_{3\mu})
\label{neu.6}
\end{eqnarray}
we get
\begin{eqnarray}&&
W_{ A_3^\mu b_0} =0,\quad W_{ A_3^\mu b_3} =- i\frac{ p_\mu}{\Lambda^{(D-4)}p^2},\quad W_{ A_3^\mu \phi_3} =0
,\quad p^\mu W_{A_3^\mu A_3^\nu} =0,
\nonumber\\&&
p^\mu W_{ A_3^\mu B^\nu}=0.
\label{neu.7}
\end{eqnarray}
From the  $U(1)$ STI (\ref{st.3})
\begin{eqnarray}
-\frac{\Lambda^{(D-4)}}{g'}\Box W_{b_0}+\frac{1}{g'}\partial^\mu J_{B^\mu} + \frac{v'}{2}J_{\phi_3} =0,
\label{neu.8}
\end{eqnarray}
where
\begin{eqnarray}
v'\equiv 
\frac{1}{\Lambda^{(D-4)}} \Gamma_{K_0},
\label{neu.9}
\end{eqnarray}
we get
\begin{eqnarray}
W_{ b_0 A_3^\mu} =0,~~ W_{b_0 B^\mu } = i\frac{ p_\mu}{\Lambda^{(D-4)}p^2},
\quad W_{b_0 \phi_3} =- \frac{v' g'}{2\Lambda^{(D-4)}p^2}
,~~  W_{ b_0 b_0 } =0, ~~ W_{  b_0 b_3} =0.
\label{neu.10}
\end{eqnarray}
From the $SU(2)$ STI (\ref{brst.13.1}) 
\begin{eqnarray}
\int d^Dx \, \Big (
- W_{ A^*_{a\mu}}  J_{a\mu}
-
 W_{ \phi_a^*} K_a
  +  \Lambda^{(D-4)}\eta_a W_{b_a}\Bigr)
 = 0 
\label{neu.11}
\end{eqnarray}
we get
\begin{eqnarray}
&&
\!\!\!\!\!\!\!\!\!\!\!\!\!\!\!\!\!\!\!\!\!\!\!\!\!\!
W_{b_3 b_0} =0,~ W_{b_3 B^\mu } = 0,
~ W_{b_3 A_3^\mu} = \frac{1}{\Lambda^{(D-4)}}W_{\bar c_3 A^*_{a\mu}}
,~  W_{ b_3 \phi_3 } = \frac{1}{\Lambda^{(D-4)}}W_{\bar c_3 \phi^*_3},
~ W_{  b_3 b_3} =0.
\label{neu.12}
\end{eqnarray}
Eqs. (\ref{neu.7}) and (\ref{neu.12}) imply the interesting result
\begin{eqnarray}
W_{\bar c_3 A^*_{a\mu}} = i\frac{ p^\mu}{p^2}.
\label{neu.13}
\end{eqnarray}
\par
We now consider the 1PI two-point functions.
From the $U(1)$ LGE
\begin{eqnarray}
\Gamma_{ b_0} =  \Lambda^{(D-4)} 
\partial^\mu B_{\mu }
\label{neu.16}
\end{eqnarray}
we get
\begin{eqnarray}&&
\G_{ b_0B^\mu} =  -i\Lambda^{(D-4)} p_\mu, \quad \G_{ b_0A_3^\mu}=0,\quad \G_{ b_0 b_3}=0
\nonumber\\&&
\G_{ b_0\phi_3}=0, \quad \G_{ b_0 b_0}=0.
\label{neu.17}
\end{eqnarray}
From the $SU(2)$ LGE (\ref{b.eq})
\begin{eqnarray}
\Gamma_{ b_a} =  \Lambda^{(D-4)} \Big ( 
 D^\mu[V](A_\mu - V_\mu)\Big )_a 
\label{neu.18}
\end{eqnarray}
we get
\begin{eqnarray}&&
\Gamma_{ b_3A_3^\mu} =  -i \Lambda^{(D-4)}p_\mu, \quad \Gamma_{ b_3B^\mu}=0,\quad \Gamma_{ b_3 b_3}=0
\nonumber\\&&
\Gamma_{ b_3\phi_3}=0, \quad \G_{ b_0 b_3}=0.
\label{neu.19}
\end{eqnarray}
From the $U(1)$ STI (\ref{st.3})
\begin{eqnarray}
- {\frac{2}{g'}}\Lambda^{(D-4)} \Box b_0
- {\frac{2}{g'}}\partial^\mu  \Gamma_{ B^\mu}
-{\frac{1}{\Lambda^{(D-4)}}}
\Gamma_{K_0} \Gamma_{\phi_3}
=0
\label{neu.20}
\end{eqnarray}
we get
\begin{eqnarray}&&
p^\mu\G_{ B^\mu \phi_3} =  -i\frac{v'g'}{2}\G_{\phi_3 \phi_3}, \quad
 p^\mu\G_{ B^\mu A_3^\nu} =  -i\frac{v'g'}{2}\G_{\phi_3 A_3^\nu}
\nonumber\\&&
p^\mu\G_{ B^\mu B^\nu} =  -i\frac{v'g'}{2}\G_{\phi_3  B^\nu}
\quad \Longrightarrow p^2 \Gamma_L^{BB}=\biggl(\frac{v'g'}{2}\biggr)^2
\G_{\phi_3 \phi_3} .
\label{neu.21}
\end{eqnarray}
From the $SU(2)$ STI (\ref{brst.13})
\begin{eqnarray}
&& \int d^Dx \,\Biggl[ \Lambda^{-(D-4)}\Big (
 \Gamma_{ A^*_{a\mu}}  \Gamma_{ A_a^\mu}
+
 \Gamma_{ \phi_a^*}  \Gamma_{ \phi_a}
\Big )
+ b_a  \Gamma_{ \bar c_a}
\Biggr] = 0
\label{neu.22}
\end{eqnarray}
we get
\begin{eqnarray}&&
p_\mu \Gamma_{ c(p) A_{3\mu}^*}= i\Gamma_{c(p)\bar c} 
\nonumber\\&&
\Gamma_{ c A_{3\mu}^*}\Gamma_{  A_3^\mu \phi_3}+
\Gamma_{c\phi_3^*}\Gamma_{\phi_3\phi_3} =0
\nonumber\\&&
\Gamma_{ c A_{3\mu}^*}\Gamma_{  A_3^\mu B^\nu}+
\Gamma_{c\phi_3^*}\Gamma_{\phi_3 B^\nu} =0
\nonumber\\&&
\Gamma_{ c A_{3\mu}^*}\Gamma_{  A_3^\mu A_3^\nu}+
\Gamma_{c\phi_3^*}\Gamma_{\phi_3 A_3^\nu} =0.
\label{neu.23}
\end{eqnarray}
From eqs. (\ref{neu.21}) and (\ref{neu.23}) we get
\begin{eqnarray}&&
\Gamma_{ c A_{3\mu}^*}= i\frac{p^\mu}{p^2} \Gamma_{c\bar c}
\nonumber\\&&
\Gamma_{  A_3^\mu \phi_3}=
i \frac{p_\mu\Gamma_{c\phi_3^*}}{\Gamma_{c\bar c}}\Gamma_{\phi_3\phi_3}
\nonumber\\&&
\Gamma_L^{AB}= ip^\nu\frac{\Gamma_{c\phi_3^*}}{\Gamma_{c\bar c}}\Gamma_{\phi_3 B^\nu}
= {-} p^2 \frac{2}{v' g'}\frac{\Gamma_{c\phi_3^*}}{\Gamma_{c\bar c}} \Gamma_L^{BB}
= {-}\frac{v' g'}{2}\frac{\Gamma_{c\phi_3^*}}{\Gamma_{c\bar c}} \Gamma_{\phi_3\phi_3}
\nonumber\\&&
\Gamma_L^{AA}= ip^\nu\frac{\Gamma_{c\phi_3^*}}{\Gamma_{c\bar c}}\Gamma_{\phi_3 A_3^\nu}
=  p^2\biggl(\frac{\Gamma_{c\phi_3^*}}{\Gamma_{c\bar c}}\biggr)^2 \Gamma_{\phi_3\phi_3}.
\label{neu.24}
\end{eqnarray}
From the condition in eq. (\ref{neu.0})  we get the following
constraints
\begin{eqnarray}
(\Gamma \cdot W)_{A^\mu\phi} =0, \quad
&\Longrightarrow&  \frac{2}{v'g'}\Gamma_L^{BA} W_{\phi\phi}= 
{\Lambda^{D-4}}
W_{b_3\phi}
\label{charge.46.1}
\\
(\Gamma \cdot W)_{B^\mu\phi} =0, \quad
&\Longrightarrow& \Gamma_{\phi\phi} W_{\phi\phi}=-1
\label{charge.46.2}
\\
(\Gamma \cdot W)_{  A  b_0 } =0, \quad
&\Longrightarrow&  \Gamma_L^{AB}- \Gamma_L^{BA}=0
\label{charge.46.3}
\\
(\Gamma \cdot W)_{  A  b_3  }=0, \quad
&\Longrightarrow&
\frac{1}{p^2\Lambda^{D-4}} \Gamma_L^{ A   A} 
+ \frac{2}{v'g'}\Gamma_L^{ A B  }W_{\phi b_3  }=0
\label{charge.46.4}
\\
(\Gamma \cdot W)_{   B   A   }
 =0, \quad
&\Longrightarrow&
\Gamma_T^{BA}  W_T^{ A   A  }+\Gamma_{T}^{BB}  W_T^{BA}=0
\label{charge.46.5}
\\
(\Gamma \cdot W)_{   A   B   }
 =0, \quad
&\Longrightarrow&
\Gamma_T^{  A    A}  W_T^{ A   B  }+\Gamma_T^{ A   B}  W_T^{B   B  }=0
\label{charge.46.6}
\\
(\Gamma \cdot W)_{   A   A   }
 =-I\!\!I, \quad
&\Longrightarrow&
\Gamma_T^{AA}  W_T^{AA}+\Gamma_T^{AB}  W_T^{BA}=-1
\label{charge.46.7}
\\
(\Gamma \cdot W)_{   B    B   }=-I\!\!I, \quad
&\Longrightarrow&
\Gamma_T^{BA}  W_T^{AB}+\Gamma_T^{BB}  W_T^{BB}=-1
\label{charge.46.8}
\end{eqnarray}
From eqs. (\ref{charge.46.1}) and  (\ref{charge.46.4}) we can deduce the
following identity
\begin{eqnarray}
\frac{\Gamma_L^{AA}}{p^2}= \biggl(\frac{2|\Gamma_L^{AB}|}{v' g'}\biggr)^2 \frac{1}{\Gamma_{\phi_3\phi_3}}.
\label{charge.48.-1}
\end{eqnarray}
Subsequently we use eq. (\ref{neu.24})
\begin{eqnarray}
\Gamma_L^{AA}=|\Gamma_L^{AB}|^2 \frac{1}{\Gamma_L^{BB}},
\label{charge.48.0}
\end{eqnarray}
i.e. the $2\times 2$ determinant
\begin{eqnarray}
\Delta_L \equiv \Gamma_L^{AA}\Gamma_L^{BB}- |\Gamma_L^{AB}|^2  =0
\label{charge.48.1}
\end{eqnarray}
and moreover again from eq. (\ref{neu.24})
\begin{eqnarray}
\frac{\Gamma_L^{AA}}{\Gamma_L^{AB}}= \frac{\Gamma_L^{BA}}{\Gamma_L^{BB}}
=-\frac{2p^2} {v'g'}\frac{\Gamma_{c_3\phi_3^*}} {\Gamma_{c_3\bar c_3}}.
\label{charge.48.2}
\end{eqnarray}


\begin{thebibliography}{99}

\bibitem{Bettinelli:2008ey}
  D.~Bettinelli, R.~Ferrari and A.~Quadri,
  Int.\ J.\ Mod.\ Phys.\  A {\bf 24} (2009) 2639
  [arXiv:0807.3882 [hep-ph]].


\bibitem{higgs}
  P.~W.~Higgs,
  Phys.\ Lett.\  {\bf 12} (1964) 132; 
  Phys.\ Rev.\ Lett.\  {\bf 13} (1964) 508;
  Phys.\ Rev.\  {\bf 145} (1966) 1156;
  F.~Englert and R.~Brout,
  Phys.\ Rev.\ Lett.\  {\bf 13} (1964) 321;
  G.~S.~Guralnik, C.~R.~Hagen and T.~W.~B.~Kibble,
  Phys.\ Rev.\ Lett.\  {\bf 13} (1964) 585;
  T.~W.~B.~Kibble,
  Phys.\ Rev.\  {\bf 155} (1967) 1554.
  

\bibitem{Ferrari:2005ii}
  R.~Ferrari,
  JHEP {\bf 0508}, 048 (2005)
  [arXiv:hep-th/0504023].

\bibitem{Bettinelli:2007tq}
  D.~Bettinelli, R.~Ferrari and A.~Quadri,
  Phys.\ Rev.\  D {\bf 77} (2008) 045021
  [arXiv:0705.2339 [hep-th]].

\bibitem{Ferrari:2004pd}
  R.~Ferrari and A.~Quadri,
  JHEP {\bf 0411} (2004) 019
  [arXiv:hep-th/0408168].


\bibitem{Bettinelli:2007zn}
  D.~Bettinelli, R.~Ferrari and A.~Quadri,
  Int.\ J.\ Mod.\ Phys.\  A {\bf 23}, 211 (2008)
  [arXiv:hep-th/0701197].


\bibitem{Ferrari:2000yp}
  R.~Ferrari, M.~Picariello and A.~Quadri,
  Annals Phys.\  {\bf 294}, 165 (2001)
  [arXiv:hep-th/0012090].

\bibitem{Veltman:1968ki}
  The nonlinear sigma model content of massive Yang-Mills theory has been
  considered by many authors. See e.g.
  M.~J.~G.~Veltman,
  Nucl.\ Phys.\  B {\bf 7} (1968) 637.


\bibitem{Bettinelli:2007kc}
  D.~Bettinelli, R.~Ferrari and A.~Quadri,
  JHEP {\bf 0703} (2007) 065
  [arXiv:hep-th/0701212].


\bibitem{Ferrari:2005va}
  R.~Ferrari and A.~Quadri,
  Int.\ J.\ Theor.\ Phys.\  {\bf 45}, 2497 (2006)
  [arXiv:hep-th/0506220].


\bibitem{Henneaux:1998hq}
  M.~Henneaux and A.~Wilch,
  Phys.\ Rev.\  D {\bf 58} (1998) 025017
  [arXiv:hep-th/9802118].


\bibitem{Bettinelli:2007cy}
  D.~Bettinelli, R.~Ferrari and A.~Quadri,
  Phys.\ Rev.\  D {\bf 77} (2008) 105012
  [arXiv:0709.0644 [hep-th]].

\bibitem{Jegerlehner:2000dz}
  F.~Jegerlehner,
  Eur.\ Phys.\ J.\  C {\bf 18} (2001) 673
  [arXiv:hep-th/0005255].


\bibitem{Froissart:1961ux}
  M.~Froissart,
  Phys.\ Rev.\  {\bf 123} (1961) 1053.

\bibitem{Lee:1977eg}
  B.~W.~Lee, C.~Quigg and H.~B.~Thacker,
  Phys.\ Rev.\  D {\bf 16}, 1519 (1977).



\bibitem{Longhitano:1980iz}
  A.~C.~Longhitano,
  Phys.\ Rev.\  D {\bf 22} (1980) 1166.

\bibitem{Longhitano:1980tm}
  A.~C.~Longhitano,
  Nucl.\ Phys.\  B {\bf 188} (1981) 118.



\end{thebibliography}
\end{document}